\documentclass[aps, pra, amsmath,showkeys, showpacs, preprintnumbers, amssymb,graphicx, notitlepage, longbibliography]{revtex4-1} 

\usepackage{mathtools}
\usepackage{amssymb}
\usepackage{amsfonts}
\usepackage{amsmath}
\usepackage{amsthm}
\usepackage{dsfont}
\usepackage{braket}
\usepackage{bm}				 

\usepackage{graphicx}
\usepackage{xcolor}

\usepackage{circuitikz}
\usepackage{qcircuit}

\usepackage{hyperref}
\usepackage[capitalise]{cleveref}

\usepackage{stmaryrd}		
\usepackage{stackengine}

\crefname{pluralequation}{eqs.}{eqs.}

\newcommand{\eg}{e.\,g.\ }
\newcommand{\ie}{i.\,e.\ }
\newcommand{\etal}{\emph{et al.\ }}

\DeclareMathOperator{\Tr}{Tr}
\renewcommand{\i}{{\mathrm{i}}}
\DeclareMathOperator*{\diffd}{d\!}

\newcommand{\e}{{\mathrm{e}}}

\theoremstyle{plain}
\newtheorem{thm}{Theorem}[section]
\newtheorem{lem}[thm]{Lemma}

\begin{document}

\title{Realizing modular quadrature measurements via a tunable photon-pressure coupling in circuit-QED}
\author{Daniel J. Weigand}
\affiliation{QuTech, Delft University of Technology, Lorentzweg 1, 2628 CJ Delft, The Netherlands}
\author{Barbara M. Terhal}
\affiliation{QuTech, Delft University of Technology, Lorentzweg 1, 2628 CJ Delft, The Netherlands}
\affiliation{Forschungszentrum J\"ulich GmbH, J\"ulich, Germany}
\date{\today}

\begin{abstract}
One of the most direct preparations of a Gottesman-Kitaev-Preskill qubit in an oscillator uses a tunable photon-pressure (also called optomechanical) coupling of the form $g \hat{q} a^{\dagger} a$, enabling to imprint the modular value of the position $\hat{q}$ of one oscillator onto the state of an ancilla oscillator. We analyze the practical feasibility of executing such modular quadrature measurements in a parametric circuit-QED realization of this coupling.
We provide estimates for the expected GKP squeezing induced by the protocol and discuss the effect of photon loss and other errors on the resulting squeezing.
\end{abstract}






\maketitle

\tableofcontents

\section{Introduction and Motivation}
Bosonic quantum error correction encoding quantum information into oscillator space(s) has gained much experimental interest in the last few years (\eg \cite{Ofek+:cat, sun+:baby-bin, fluehmann:GKP, CI+:grid, grimm+:kerrcat}). 
A good reason to use a single oscillator instead of multiple qubits to encode quantum information redundantly is that control, manipulation and fabrication of a single oscillator can be easier than that of multiple oscillators or qubits.
In other words, bosonic error correction can be a hardware-efficient way~\cite{leghtas:cat} of producing novel qubits which hopefully have longer coherence versus gate times than current popular members of the qubit family, such as the transmon qubit in superconducting devices.
A promising code encoding a qubit into a single oscillator is the so-called GKP code after the proposal by Gottesman, Kitaev and Preskill in 2001~\cite{GKP}.
This code has the ability to correct small shifts in phase space, but has also been shown to be very competitive, as compared to other code contenders, with respect to photon loss errors~\cite{Albert+:bos_codes,Noh.etal.2018:quant-ph}
For an encoded qubit such as the GKP qubit, important aspects of its performance will be determined by the ability to reliably prepare or measure the qubit in the $Z$ and $X$-basis, perform single- and two-qubit gates on it (CNOT, Hadamard and T gates), as well as execute quantum error correction in a fault-tolerant manner. Theoretical methods and circuits to obtain these components have been discussed, for example, in~\cite{GKP,GK:FT, TW:GKP} and~\cite{SCC:GKP}.

In particular, as the GKP qubit states are highly non-classical `grid' states, one can ask about the best method to prepare such states from the vacuum, given a coupling with an ancilla system which is subsequently measured.
The original GKP paper~\cite{GKP} briefly suggested that a photon-pressure coupling between the target oscillator, --in which the state is to be prepared--, and an ancilla oscillator would be useful in this respect. Through such an interaction the ancilla oscillator acquires a frequency shift which depends on the quadrature $\hat{q}_T=\frac{1}{\sqrt{2}}(a_T+a_T^{\dagger})$ of the target oscillator $T$.
Instead of measuring this frequency shift, the aim is then to measure just the effective rotation that it induces on an initial state in the ancilla oscillator after a specific interaction time.
Values for $q$ which differ in the ancilla oscillator state being rotated by a full period are thus not distinguished.
This means that the interaction can be used to realize modular measurements of $\hat{q}$ and $\hat{p}$. Such modular quadrature measurements commute when the product of the moduli is a multiple of $2\pi$.
 It is precisely these modular quadrature measurements which are required to prepare a GKP qubit: they  can also be used to stabilize a GKP qubit~\cite{CI+:grid} or perform quantum error correction.

Modular quadrature measurements~\cite{fluehmann:modular} are of fundamental interest since commuting quadrature measurements allow one to measure both quadratures without fundamental Heisenberg uncertainty, with possible application in displacement sensing in the microwave domain~\cite{DTW:sensor}.
The use of such modular variables directly gives rise to a mixed position-momentum representation of a state in phase space: Zak first formulated this idea, giving a mixed momentum-position state of electrons in solids, see the review~\cite{englert:zak} and references therein.

In this paper we present a circuit-QED set-up for coupling two (close to harmonic) oscillators via a tunable photon-pressure coupling with the aim of realizing a modular quadrature measurement in one of the oscillators, see \cref{sec:setup-global}.
This measurement requires a full measurement of the ancilla oscillator state, which in circuit-QED can be obtained by releasing this state, via a lossy oscillator, to a transmission line where the signal gets amplified and finally read-out at room temperature.
In \cref{sec:release} we briefly discuss previous work on such release or `switch' mechanisms which can be turned on and off to high approximation. Prior to this, we provide an overview of our modular quadrature measurement scheme in \cref{sec:simple}. Other and related means to obtain a photon-pressure coupling in circuit-QED are reviewed in \cref{sec:previous}.

In \cref{sec:meas} we estimate the expected performance of the modular quadrature measurement: this is expressed in terms of how `squeezed' a GKP qubit can be obtained through this measurement.
The squeezing effectively captures how much one becomes an eigenstate of the operator which is measured.
The aim here is to do a strong modular quadrature measurement, unlike some of the previous work~\cite{TW:GKP, fluehmann:GKP, CI+:grid} in which the measurement is built up from a sequence of weak measurements via coupling to ancilla qubits.
In the latter approach the strong measurement, --which is effectively a phase estimation or eigenvalue measurement of a unitary displacement operator--, is obtained through ancilla qubit measurements, each contributing at most 1 bit of phase information.
The strength of the modular quadrature measurement proposed in this paper will be governed by the number of photons in the ancilla oscillator used to perform the measurement: the more photons are in the ancilla oscillator, the stronger the measurement which is realized.
We will compare our new proposal with the sequential qubit measurement scheme~\cite{TW:GKP} using a transmon qubit~\cite{CI+:grid} or Kerr-cat qubit~\cite{puri+:detector-GKP, grimm+:kerrcat} in \cref{sec:compare}, also with respect to error feedback to the target oscillator.
As the preparation protocol will inevitably suffer from imperfections, we discuss several noise mechanisms and their effect in \cref{sec:noise}. We end the paper with a conclusion and a discussion, summarizing our findings, in \cref{sec:discuss}.

\subsection{Preliminaries}
This section collects a few conventions and the definition of the GKP code. We use $\hat{q}=\frac{1}{\sqrt{2}} (a+a^{\dagger})$ and $\hat{p}=\frac{\i}{\sqrt{2}}(a^{\dagger}-a)$ so that $[\hat{q},\hat{p}]=\i I$\footnote{In some texts the quadrature operators are defined as $X=\frac{1}{2}(a+a^{\dagger})$ and $P=\frac{\i}{2}(a^{\dagger}-a)$ instead, see \eg\cite{book:HR}.
The latter convention has the advantage of connecting directly to the real and imaginary part of a coherent state $\ket{\alpha}$, while our choice is used by~\cite{GKP} so we adhere to this convention.} Phase space displacements (translations) are denoted, in standard form, as $D(\alpha)=\exp(\alpha a^{\dagger}-\alpha^* a)$.

The (square) GKP code is defined by two commuting code stabilizers equal to $S_q=\exp(\i 2 \sqrt{\pi} \hat{q})$ and $S_p=\exp(-\i 2 \sqrt{\pi}\hat{p})$. These operators act as shift or displacement operators in phase-space, that is $S_q \ket{p}=\ket{p+2\sqrt{\pi}}$ and $S_p\ket{q}=\ket{q+2\sqrt{\pi}}$. States which have eigenvalue 1 with respect to these operators are thus invariant under these translations in phase space.
There are two operators $X=\exp(-\i \sqrt{\pi} \hat{p})$ and $Z=\exp(\i \sqrt{\pi} \hat{q})$ which both commute with $S_p$ and $S_q$, while $XZ =-ZX$ and hence they are the logical Pauli operators, --equal to half-stabilizer shifts--, of the encoded qubit. Note that the operators $S_p,S_q,Z$ and $X$ only square to the identity in the codespace.
Measuring the eigenvalue of a unitary operator such as $S_q$ is equivalent to measuring the value for $q$ \emph{modulo} $\sqrt{\pi}$, as all values $\hat{q}=q_{\rm meas}+k \sqrt{\pi}$ for $k \in \mathbb{Z}$ give the same eigenvalue $\exp(\i 2\sqrt{\pi} q_{\rm meas})$ for $S_q$. Said differently, a modular quadrature measurement is the measurement of the eigenvalue of a unitary displacement operator.

Since the eigenvalue of a unitary operator is a phase, the phase variance of the post-measurement state captures how precisely the eigenvalue is measured. This phase variance or uncertainty is expressed by effective squeezing parameters, one for the measurement of $S_p$, and one for the measurement of $S_q$. These squeezing parameters can be chosen (see details and relation with Holevo phase and regular quadrature variance in~\cite{DTW:sensor}) as
\begin{align}
	&\Delta_{p}=\Delta_p(\rho)=\sqrt{\frac{1}{2\pi} \ln\left(\frac{1}{|{\rm Tr} S_{p} \rho|^2}\right)}, &&\Delta_{q}=\Delta_q(\rho)=\sqrt{\frac{1}{2\pi} \ln\left(\frac{1}{|{\rm Tr} S_{q} \rho|^2}\right)}.
	\label{def:squeezing}
\end{align}
To get some intuition, note $0 \leq |{\rm Tr} S_q \rho|\leq 1$ in general, but will be 1 if $\rho$ is an eigenstate with a particular eigenvalue for $S_q$ and $0$ if it is a uniform superposition of eigenstates, hence $|{\rm Tr} S_q \rho|$ expresses the \emph{sharpness} or concentration of $\rho$ around an $S_q$ eigenstate.
Classically, the topic of circular statistics is well-established, see \eg\cite{book:Jammalamadaka.Sengupta:CircularStatistics}: for a probability distribution $\mathbb{P}(\theta)$ over an angle $\theta \in [0,2\pi)$, the circular standard deviation is defined as $\sqrt{\ln{( 1/{|\int \diffd\theta\; \mathbb{P}(\theta) \exp(\i \theta)|}^2)}}$.
The squeezing parameters in \cref{def:squeezing} are thus a direct application of the circular standard deviation.
With the convention in \cref{def:squeezing}, the vacuum state has $\Delta_p=\Delta_q=1$ showing that it is not squeezed.
A $\Delta$-squeezed vacuum state (in $q$) has variance $\bra{\rm sq. vac.}{(q-\langle q \rangle)}^2\ket{\rm sq. vac}=\Delta^2 \Delta^2_{\rm vac}$ with $\Delta=\Delta_q$ and $\Delta_{\rm vac}=\frac{1}{2}$\footnote{We remark that Ref.~\cite{CI+:grid} uses a standard deviation $\sigma$ as the absolute standard deviation of a squeezed peak while the $\Delta$ parameter is the relative enhancement of the standard deviation as compared to the vacuum state.
This implies that we have the correspondence $\sigma^2=\Delta^2/2$ since the vacuum has variance $1/2$ by definition.}. For a Gaussian model wavefunction of an approximate GKP state it holds that $\bar{n} \approx \frac{1}{2\Delta^2}-\frac{1}{2}$~\cite{GKP, TW:GKP}. In this model an approximate GKP state equals
\begin{align*}
&\ket{\tilde{\psi}} = \frak{E} \ket{\overline{\psi}},
&&\frak{E}=\frac{1}{\sqrt{\pi \Delta^2}} \int_{\mathbb{R}^2} \diffd u \diffd v\; \exp(-(u^2+v^2)/2\Delta^2) \exp(-\i u \hat{p}+\i v \hat{q}),
\end{align*}
where $\ket{\overline{\psi}}$ is perfect GKP code state, \ie a $+1$ eigenstate of $S_p$ and $S_q$.

\subsection{Overview of Measurement Protocol}\label{sec:simple}

\begin{figure}[htb]
\includegraphics{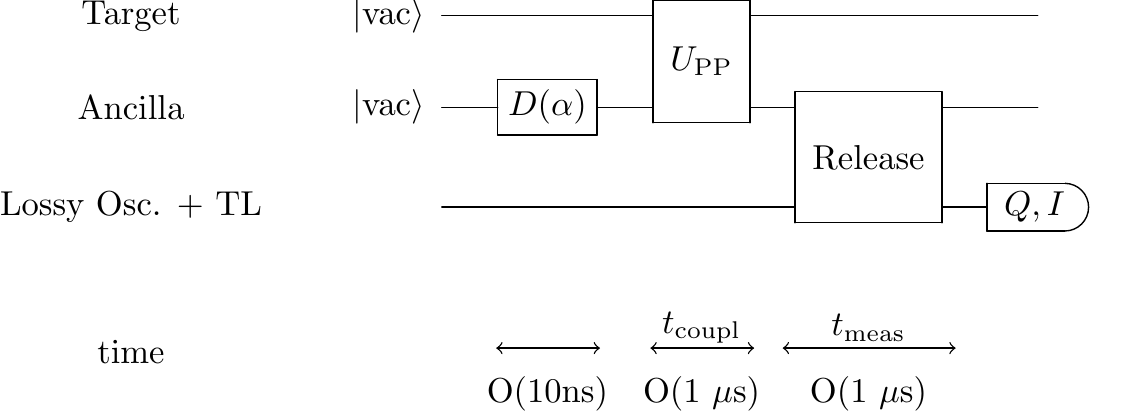}
	\caption{Timeline of the measurement protocol. First, the ancilla oscillator is initialized to a coherent state $\ket{\alpha}$. Then, the parametric drive is turned on for time $t_{\rm coupl}$, coupling the target and ancilla oscillators with the unitary $U_{\rm PP}$. Finally, the parametric drive is turned off and the ancilla oscillator is coupled to a lossy oscillator. From this lossy oscillator, the state is released into a transmission line, where it is amplified and measured.}
\label{fig:seq}
\end{figure}

We will refer to the oscillators as \emph{target} and \emph{ancilla} oscillators, with resonance frequencies $\omega_T$ resp.\ $\omega_A$ and $\omega_A \gg \omega_T$. We will use $a, a^\dag$ (resp.\ $b,b^\dag$) as annihilation and creation operators of the ancilla (resp.\ target oscillator).
Targeted values of coupling strengths and oscillator decay rates are summarized in \cref{tab:properties}.
The aim is to describe a set-up allowing for the measurement of both stabilizers $S_p, S_q$ and/or the logical shifts $X,Z$. For example, one can prepare a GKP grid state in the target oscillator from the vacuum by performing a modular measurement of both $\hat{p}$ and $\hat{q}$, \ie measure $S_p$ and $Z$ in sequence.

The sequence of events to enact a single modular quadrature measurement, say $S_q$, is shown in Fig.~\ref{fig:seq}. We start both oscillators in the vacuum state.
First we create a coherent state $\ket{\alpha}$ in the ancilla oscillator by driving this oscillator with a short $O(10){\rm ns}$ pulse. Now we turn on a strong photon-pressure coupling between target and ancilla oscillator for time $t_{\rm coupl}$: we discuss this in detail in \cref{sec:circuit}. In the rotating frame of both oscillators (ancilla oscillator at $\omega_A$ and target oscillator at $\omega_T$) we thus turn on the Hamiltonian
\begin{align}
	H_{\rm PP} = g a^\dag a (b^\dag + b)=\sqrt{2} g a^{\dagger} a \hat{q}_T,
	\label{eq:H_desired}
\end{align}
for some time $t_{\rm coupl}$.
Here, and throughout the rest of this paper we use the convention $\hbar = 1$.
We note that the fact that this Hamiltonian is time-independent in the rotating frame of target oscillator is non-trivial: it requires a parametric drive by a classical field, \ie a pump or a flux-drive to accomplish this. By changing the phase of this classical field we can change the coupling to be proportional to $a^{\dagger} a \,\hat{p}_T$ enabling to perform a modular measurement of $\hat{p}$ (or another rotated quadrature).

If the interaction in \cref{eq:H_desired} is turned on for a time $t_{\rm coupl}=\sqrt{2\pi}/g$, it implements the following unitary between target and ancilla oscillator
\begin{align}
	\label{eq:U_{TA}}
	U_{\rm PP} = \exp(\i 2\sqrt{\pi}\hat{q}_T a^\dag a) = S_{q_T}^{a^\dag a},
\end{align}
where \(S_{q_T}\) is a stabilizer of the GKP code acting on the target oscillator.
From now on, we will drop the subscript $T$ in the stabilizers $S_{q_T}, S_{p_T}$ and logicals $Z_T, X_T$ as all these operators always act on the target oscillator.
It follows that the coherent state $\ket{\alpha}$ in the ancilla oscillator rotates by an amount which depends on the eigenvalue phase of $S_{q}$.
Thus measuring the angle over which the state $\ket{\alpha}$ rotates corresponds to measuring the eigenvalues of $S_{q}$. A coherent state $\ket{\alpha}$ naturally has an angle uncertainty which gets larger with smaller amplitude $|\alpha|$, putting an $\alpha$-dependent bound on the accuracy with which one can project onto an eigenstate of $S_{q}$.
Clearly, the larger the coherent amplitude is, the better one can resolve its phase and thus the more bits of information one gets about the eigenvalue phase of the measured operator $S_q$\footnote{Note that it would defeat the purpose to get the fullest possible angle information if we were to subsequently map the ancilla oscillator state to a state of a qubit as in~\cite{blumoff+:parity,NG:parity}: the ancilla qubit carries at most one bit of information}.

After the photon-pressure interaction is turned off and the oscillators no longer interact, the state of the ancilla oscillator has to be converted to a traveling signal so that the quadratures of the rotated coherent state can be read out via the standard `heterodyne' measurement chain~\cite{eichler+:hetero}, allowing one to determine the phase of the coherent state.
We do not claim any original contribution for such a release mechanism, but discuss known previous work in \cref{sec:release}.

\begin{figure}[htb]
	\includegraphics[width=\textwidth]{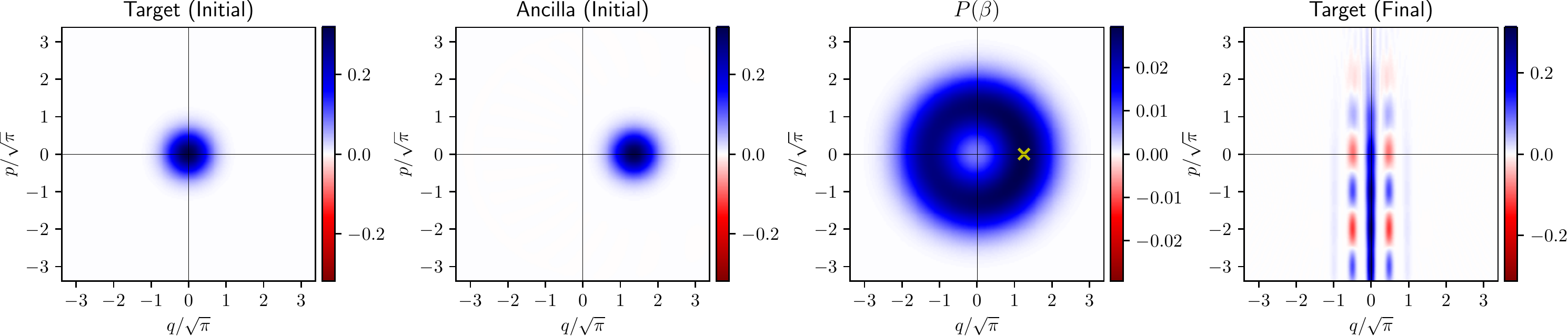}\\
	\ \\ 
	\includegraphics[width=\textwidth]{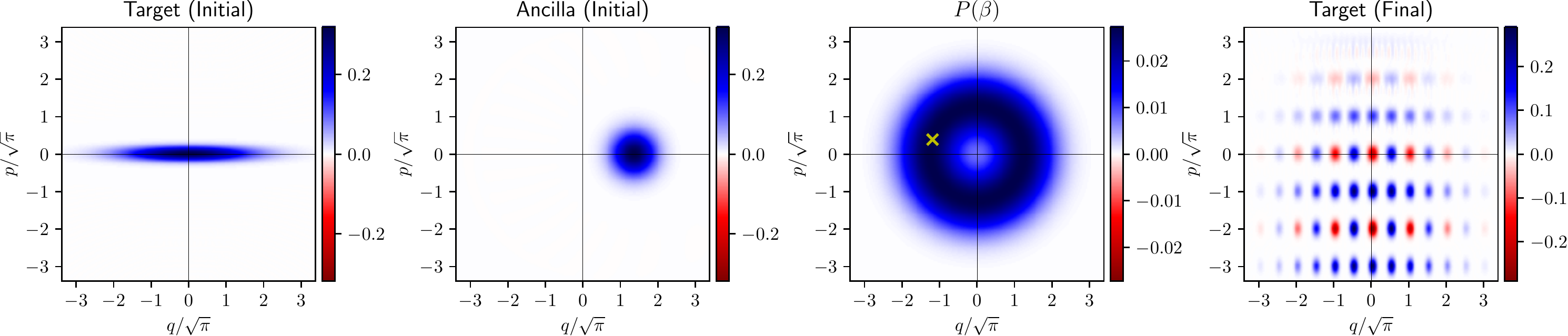}
	\caption{
		Wigner functions of states in target and ancilla oscillator and probability distribution $\mathbb{P}(\beta)$ over measurement results of heterodyne measurement of the ancilla oscillator mode.
		The initial state of the ancilla oscillator is the coherent state $\ket{\alpha=\sqrt{3}}$.
		The measurement result is the one with the maximum likelihood with respect to $\mathbb{P}(\beta)$ (marked by yellow cross).
		Top row: Starting with a vacuum state ($\Delta_q=\Delta_p=1$) in the target mode, a measurement of $S_q$ results in an effective squeezing of the final state of $\Delta_q=0.18$, while $\Delta_p=1$ is unchanged. The final state is most like the GKP $\ket{-}$ state for the following reason.
		We start with a vacuum state --- which is closest to the $+1$ eigenstate of $X$. Besides, the measurement result gives an eigenvalue of $S_q$ close to $+1$ so we are in the GKP code space.
		In order to center the outgoing state we apply an additional unconditional displacement equal to $Z^{-3}$ which changes the initial eigenvalue $+1$ of $X$ to $-1$.
		Bottom row: the initial state in the target mode is a squeezed vacuum state with $\Delta_q=3$ and $\Delta_p=1/3$. The effective squeezing of the final state $\Delta_p=1/3$ is again unchanged, while $\Delta_q=0.18$ for the outgoing state. The resulting state is squeezed with respect to both quadratures.
		Now the final state is close to a GKP $\ket{-}$ displaced by half a logical, \ie $X^{-1/2}$, for the following reason.
		Again, we started with a squeezed vacuum state, which is closest to the $+1$ eigenstate of $X$ and the unconditional displacement is $Z^{-3}$, which changes the eigenvalue to $-1$.
		However, the measurement result now gives an eigenvalue of $S_q$ close to $-1$, indicating that the state is shifted out of the code space, by half a logical $X$.}
\label{fig:wigner-simple}
\end{figure}

In \cref{sec:toy} we formally model the effect of the whole measurement protocol: in \cref{fig:wigner-simple} we show the effect of the protocol using a coherent state with mean photon number $\bar{n}=3$. If we integrate the Wigner function of the outgoing state over the $p$-coordinate, we obtain the probability distribution over $q$ which is clearly peaked, with periodicity $2\sqrt{\pi}$.

Note that the support of these peaks lies within the uncertainty of the original vacuum state:
The measurement of $S_q$ does not enlarge the $q$-support of the input wave-function, it only applies a modular structure to it.
The measurement of $S_q$ \emph{does} enlarge the $p$-support of the input wave-function as is visible from the Wigner function of the outgoing state. Thus, if we were to follow the measurement of $S_q$ by a measurement of $S_p$, we would obtain the signature grid-like Wigner function of an approximate GKP state.
Alternatively, we start with a squeezed state (squeezing in $p$) and only measure $S_q$, see the bottom row in \cref{fig:wigner-simple} to obtain a grid-like GKP Wigner function.

\subsubsection{GKP Qubit Readout}
The preparation of a GKP grid state should also be accompanied by a mechanism for measuring the GKP qubit in the $Z$ or $X$-basis. A useful fault-tolerant $Z$-measurement is a measurement in which the quadrature $q$ is measured so that finding the quadrature $q$ closer to an even (resp.\ odd) multiple of $\sqrt{\pi}$ leads to inferring the state 0 (resp.\ 1).
A simple method is to use the photon-pressure coupling and replace $S_q$ by the logical operator $Z$ to nondestructively measure $Z$. If $t_{\rm coupl}$ is turned on for half the time, such that the ancilla oscillator is either not rotated ($Z \approx 1$) or rotated by $\pi$ ($Z \approx -1$), then subsequent release and measurement of the state of the ancilla oscillator reveals the eigenvalue of $Z$. Readout of the Pauli $X$ operator could proceed analogously.


\subsubsection{Why probing the ancilla oscillator's frequency reveals the wrong information}

Our scheme is demanding in requiring a high-Q ancilla oscillator (low $\kappa_c$) whose state should be measured through a tunable release or switch mechanism (switching to higher $\kappa_{\rm open}$) followed by a circuit-QED heterodyne measurement. The photon-pressure coupling induces a frequency shift in the ancilla oscillator which depends on the quadrature of the target oscillator.
We could imagine measuring such a frequency shift by probing the ancilla oscillator with a microwave tone as is done in the standard dispersive measurement in circuit-QED~\cite{blais:meas}, without switching the effective decay rate of the ancilla oscillator from low to high for state release. Here we briefly comment on the fact that this method will not work as we will obtain direct rather than modular information about the target oscillator quadrature $q_T$.

Imagine we would weakly apply a microwave drive to the ancilla oscillator (decay rate $\kappa$) at some frequency $\omega$, starting at some initial time $t=0$. Also at time $t=0$, we have turned on the photon-pressure coupling so that the resulting Hamiltonian of ancilla and target oscillator is $H_{\rm PP}=(\omega_A +g \hat{q}_T) a^{\dagger} a$ in the rotating frame of the target oscillator at angular frequency $\omega_T$.
We can thus view the photon-pressure coupling as an effective change in the resonant frequency of the ancilla oscillator, which leads to a phase change of the outgoing field as compared to the incoming field. For simplicity, we take the weak drive to be modeled by a plane-wave input field $b_{\rm in}[\omega]$ at frequency $\omega$. The input-output formalism (see \eg\cite{book:WM, BDT:meas}) gives the phase of the reflected output field as
\begin{align}
&b_{\rm out}[\omega]=\exp(i\varphi(\hat{q}_T, \omega)) b_{\rm in}[\omega], &&\exp(i\varphi(\hat{q}_T,\omega))=\frac{\kappa/2+i(\omega_A+g\hat{q}_T-\omega)}{\kappa/2-i(\omega_A+g\hat{q}_T-\omega)}.  \notag
\end{align}
If we choose $\omega=\omega_A$, \ie drive the ancilla oscillator at its resonant frequency, then we see that the phaseshift $\varphi(\hat{q}_T,\omega_A)$ goes from $-\pi$ at large negative $q_T$ to $\pi$ at large positive $q_T$ and displays no periodicity in $q_T$ since $\varphi=2 \arctan(2g \hat{q}_T/\kappa)$.
These considerations imply that the modular measurement of $q$ should take place in a very non-steady state regime where the ancilla resonator is first excited to create the state $\ket{\alpha}$ and decay of this state should be strongly suppressed during the photon-pressure interaction, as this decay will leak information about $q_T$. We discuss the effect of photon loss in the ancilla oscillator during the interaction in \cref{sec:loss-ancilla-int}.

\section{Circuit-QED Setup}
\label{sec:setup-global}
In this section, we discuss how a modular quadrature measurement can be realized. We start with a short review of previous work that realizes a photon-pressure or longitudinal coupling.
We then introduce and analyze an electric circuit that achieves strong coupling.
Finally, we discuss how the state in the ancilla oscillator can be released into a transmission line for readout.

\subsection{Previous Circuit-QED Work on Photon-Pressure and Longitudinal Coupling}
\label{sec:previous}
When the $a$-mode of a photon-pressure coupling of the form $\hat{q} a^{\dagger} a$ is very anharmonic and is used to represent a qubit, the photon-pressure coupling can be recognized as a longitudinal coupling $\hat{q} (I-Z)/2$ with Pauli $Z$ of the qubit. In this incarnation the qubit induces a state-dependent displacement on the target oscillator which can be used for (improved) qubit read-out~\cite{touzard:displace, Ikonen:meas,DBB:fast-para}.
Note that in such settings the roles of ancilla and target are reversed as compared to the setting of the GKP code, \ie the target oscillator is used for information gathering about the qubit instead of the target oscillator being used to store a GKP state.

In optomechanical systems the coupling $\hat{q} a^{\dagger} a$, with $\hat{q}$ the position of the mechanical oscillator and $a$ the annihilation operator of an optical cavity field, is arrived at naturally.
In the rotating frame of these oscillators, this coupling averages out without further time-dependent driving.
In a linearized regime where one expands around a driven optical field $\langle a\rangle=\alpha(t)$, the coupling can be used to generate an effective beam-splitter interaction with a strength depending on $|\alpha|^2$~\cite{AKM:RMP,Eichler.Petta.2017:CqedOptomechanicsExp}.
Although there has been a wide range of experimental setups and studies, the so-called single-photon coupling regime, $g \gg \kappa_A, \kappa_T$, \ie the bare coupling strength exceeds the photon loss rate of both oscillators has so far not been achieved~\cite{AKM:RMP}.
The difficulty is that in a traditional optomechanical setting, the loss rate of the optical oscillator is relatively large, while the mechanical oscillator, being low in frequency, is susceptible to thermal excitations. Thus, working with two oscillators both at some middling frequency (GHz range) can resolve this conundrum.

A good candidate to achieve a single-photon coupling at microwave frequencies is the so-called simulated optomechanical coupling, where a SQUID loop is used to couple two oscillators.
The coupling of two co-planar waveguide resonators via a SQUID loop has been analyzed by Johansson \etal\cite{Johansson.etal.2014:CqedOptomechanicsTheo}, a lumped element circuit has been implemented by Eichler and Petta~\cite{Eichler.Petta.2017:CqedOptomechanicsExp}.

We note that the experimental coupling achieved in~\cite{Eichler.Petta.2017:CqedOptomechanicsExp} is not in the so-called single photon regime, \ie the photon loss rate of the ancilla oscillator is larger than the coupling strength, $\kappa \geq g$.  It will be necessary to be in this regime for our use of this coupling.

\subsection{Circuit Analysis and Approximations}
\label{sec:circuit}
\begin{figure}[htb]
	\includegraphics{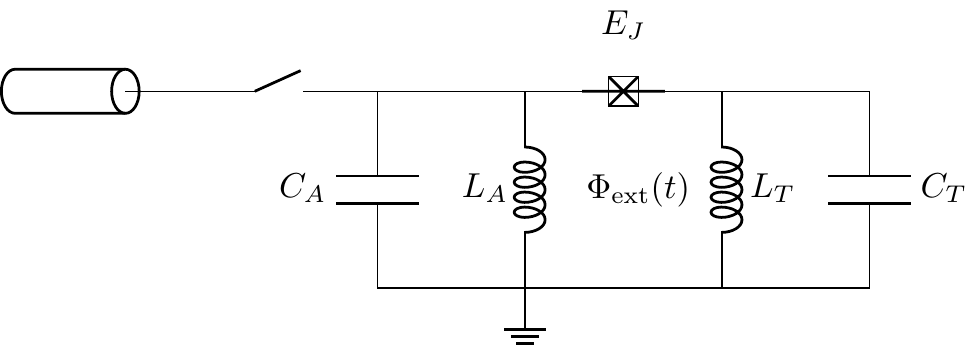}
	\caption{Electric circuit for realizing photon-pressure coupling. The target oscillator (label $T$, right) is coupled to the ancilla oscillator (label $A$, middle) via a Josephson junction.
	The coupling between the ancilla oscillator and the readout line is tunable, and only turned on during readout.
	The loop formed by the Josephson junction and the inductances $L_A, L_T$ is threaded by an external flux $\Phi_{\rm ext}(t)$, which is a classical, time-dependent variable.}
	\label{fig:circuit2}
\end{figure}

\begin{table}[htb]
	\begin{tabular}{l|c|c}
		& Ancilla Oscillator & Target Oscillator \\
\hline
Resonance frequency $f_0$ (GHz) & 10 & 0.5\\
Frequency range $f_{\rm max}-f_{\rm min}$ (MHz) & 500 & 5--10\\
Capacitance $C_m$ (pF) &0.1--1& 50--1000\\
Charging Energy $E_{C_m}/(2\pi)$ (MHz) & 20--200 & 0.02--0.4\\
Inductance $L_m$ (nH) & 0.2--3 & 0.2--3\\
Inductive Energy $E_{L_m}/(2\pi)$ (GHz) & 50--400 & 50--400\\
Third-order nonlinearity ($\sim q^3$) / g & {\rm negl.} &$ 10^{-3}$--$10^{-2}$\\
Self-Kerr ($\sim {(a^\dag a)}^2, {(b^\dag b)}^2$) / g &50\%--100\% & $10^{-3}$--$10^{-2}$\\
Targeted oscillator life time $1/\kappa$& $100\mu$s (closed), $1\mu$s (open)  & $100\mu$s \\
\hline
$E_J/(2\pi)$ (GHz) & \multicolumn{2}{c}{5--40} \\
Capacitance $C_J$ &\multicolumn{2}{c}{$C_J \ll C_A$}\\
Photon-pressure coupling $g/(2\pi)$ (MHz) & \multicolumn{2}{c}{3--15}\\
Cross-Kerr ($\sim a^\dag a b^\dag b$)/ g & \multicolumn{2}{c}{2\%--5\%} \\
Interaction time $t_{\rm coupl}$ ($\mu s$) & \multicolumn{2}{c}{0.2--1}\\
\end{tabular}
	\caption{
		Targeted properties of the two oscillators and strength of the photon-pressure coupling $g$ and error terms. 		The resonance frequency of the oscillators is dependent on the external flux, it is maximal for $x_{\rm ext}=\pi$ and minimal for $x_{\rm ext}=0$.
		All frequencies except the frequency range $f_{\rm max}-f_{\rm min}$ are given as mean values, \ie for $x_{\rm ext}=\pi/2$.
		The capacitance $C_J$ of the Josephson junction is not particularly important, as long as it is small compared to the capacitance of the ancilla oscillator $C_A$, which is the case \eg for the junction designs of the transmon and charge qubits.
		The photon-pressure coupling strength $g$ is obtained by fixing the resonance frequency, inductance and Josephson energy for the two oscillators, choosing the capacitance accordingly and using \cref{eq:H_eff}.
		The capacitance of the Josephson junction was neglected, because it is several orders of magnitude smaller than the capacitance of both oscillators.
		The nonlinear terms (third-order, self- and cross-Kerr) are given as a fraction of the coupling strength g, because they are only relevant while the drive is on.
		Note that the listed values of the self- and cross-Kerr terms are the maximal values in time (not the echoed-out values).
		In our modeling it is assumed that all losses on the ancilla oscillator are due to coupling to the transmission line.
		We denote the closed decay rate of the ancilla oscillator as $\kappa_c$ and the open decay rate as $\kappa_{\rm open}$, while the decay rate of the target oscillator is denoted as $\kappa_T$. This choice of parameters meets the condition $\kappa_c t_{\rm coupl} |\alpha|^2 \ll 1$, ensuring low photon loss during the modular quadrature measurements, easily for $\bar{n} \approx 2-4$.
		}
	\label{tab:properties}
\end{table}

To achieve the desired photon-pressure coupling, we start with the electric circuit shown in \cref{fig:circuit2}, neglecting the switchable coupling to the external world.
The GKP state will be encoded in the, low-frequency, target oscillator, shown on the very right of the figure. It is coupled via a Josephson junction to an ancilla oscillator shown in the middle.
The two oscillators are almost harmonic, with the parameters chosen such that the \emph{instantaneous} potential of the ancilla oscillator depends on the current state of the target oscillator while the potential of the target oscillator is unchanged.
This change of potential changes the resonance frequency depending on the state of the target oscillator, enabling the photonpressure coupling.
The concept is similar to the approach used by Johansson \etal, where the effective length of the ancilla slit line resonator depends on the state of the target oscillator~\cite{Johansson.etal.2014:CqedOptomechanicsTheo}.

After the interaction between ancilla and target is turned off, we envision that a coupling between transmission line and ancilla oscillator is turned on, enabling fast readout.
We note that this electric circuit has also been analyzed (operated in a different regime with very different parameters as compared to those in \cref{tab:properties}) in~\cite{Vrajitoarea.etal.2018}, with the aim to control individual Fock states as a qubit.

We envision that this circuit is realized as a superconducting lumped element circuit, using, for example, plate capacitances for getting a large $C_A$ and $C_T$, and wire structures made of superconducting material as inductance (similar to the circuit in~\cite{Eichler.Petta.2017:CqedOptomechanicsExp}).
The Lagrangian of the circuit in terms of node flux variables and their time-derivatives is
\begin{align}
\mathcal{L} = \frac{C_A \dot{\mathbf{\Phi}}_A^2}{2}  + \frac{C_T\dot{\mathbf{\Phi}}^2_T}{2}
+\frac{C_J}{2}{\left(\dot{\mathbf{\Phi}}_A-\dot{\mathbf{\Phi}}_T \right)}^2
-\frac{\mathbf{\Phi}_A^2}{2 L_A}
-\frac{\mathbf{\Phi}_T^2}{2 L_T}
+E_J \cos\left(\frac{2\pi}{\Phi_0}(\mathbf{\Phi}_T -\mathbf{\Phi}_A-\Phi_{\rm ext}(t)) \right) \notag.
\end{align}
Here $\Phi_0$ is the magnetic flux quantum and $\Phi_{\rm ext}(t)$ is a classical, time-dependent flux due to an external field. In order to obtain the Hamiltonian and define conjugate charge variables, one inverts the capacitance matrix, arriving at
\begin{align}
	H = \frac{1}{2}\frac{C_T \mathbf{Q}_A^2 + C_A \mathbf{Q}^2_T + C_J {\left(\mathbf{Q}_A+\mathbf{Q}_T\right)}^2}{C_J (C_T+ C_A) + C_A C_T}
	+\frac{\mathbf{\Phi}_A^2}{2 L_A}
	+\frac{\mathbf{\Phi}_T^2}{2 L_T}
	-E_J \cos\left(\frac{2\pi}{\Phi_0}(\mathbf{\Phi}_T -\mathbf{\Phi}_A-\Phi_{\rm ext}(t)) \right) \notag.
\end{align}
When we quantize this Hamiltonian, we have conjugate-variable commutation relations $[\mathbf{\Phi}_i,\mathbf{Q}_j] = \i \delta_{ij}$ (with $i,j=T,A$) between the flux and charge variables of the target and ancilla systems. Both flux and charge operators have eigenvalues in $\mathbb{R}$.

In the following we use that the capacitances of both oscillators are much larger than the capacitance of the Josephson junction, \ie $C_T,C_A \gg C_J$.
Up to first order in $C_J$, the Hamiltonian is then given by:
\begin{align}
	H = \frac{\mathbf{Q}_A^2}{2 C_A}  + \frac{\mathbf{Q}^2_T}{2 C_T}
	-\frac{C_J}{2}{\left(\frac{\mathbf{Q}_A}{C_A}-\frac{\mathbf{Q}_T}{C_T} \right)}^2
	+\frac{\mathbf{\Phi}_A^2}{2 L_A}
	+\frac{\mathbf{\Phi}_T^2}{2 L_T}
	-E_J \cos\left(\frac{2\pi}{\Phi_0}(\mathbf{\Phi}_T -\mathbf{\Phi}_A-\Phi_{\rm ext}(t)) \right)  \notag
\end{align}

To simplify notation, we define dimensionless conjugate variables
$\mathbf{x}_i = \frac{2\pi\mathbf{\Phi}_i}{\Phi_0}$, $\mathbf{p}_i = \frac{\Phi_0\mathbf{Q}_i}{2\pi}$, with $[\mathbf{x}_i,\mathbf{p}_i] = \i \delta_{ij}$ and a dimensionless variable
\begin{equation}
x_{\rm ext}(t)=\frac{2\pi\mathbf{\Phi}_{\rm ext}(t)}{\Phi_0}, \notag
\end{equation}
for the flux drive. We also define the charging energies $E_{C_m} = \frac{e^2}{2 C_m}$
and inductive energies $E_{L_m} = \frac{1}{4 e^2 L_m}$ for $m=T,A$, where $e$ is the elementary charge, so that
\begin{align}
	H = 4E_{C_A}\left(1-\frac{E_{C_A}}{E_{C_J}}\right)\mathbf{p}_A^2  + 4E_{C_T}\left(1-\frac{E_{C_T}}{E_{C_J}}\right)\mathbf{p}_T^2
	+\frac{8 E_{C_A}E_{C_T}}{E_{C_J}} \mathbf{p}_A \mathbf{p}_T+ U(\mathbf{x}_A,\mathbf{x}_T),
	\label{eq:H}
\end{align}
with
\begin{equation}
U(\mathbf{x}_A,\mathbf{x}_T)=\frac{E_{L_A}\mathbf{x}^2_A}{2}
	+\frac{E_{L_T}\mathbf{x}^2_T}{2}
	-E_J \cos\left(\mathbf{x}_T -\mathbf{x}_A-x_{\rm ext}(t)\right).
	\label{eq:pot}
\end{equation}
We note that the effect of the time-dependent flux-drive $x_{\rm ext}(t)$ can also be realized with a microwave drive, see details in \cref{sec:microwave_drive}.

Because we envision inductive and capacitive parameters, see \cref{tab:properties}, such that the charging energies $E_{C_A} \ll E_{L_A}$ and $E_{C_T}\ll E_{L_T}$, both $\mathbf{x}_A, \mathbf{x}_T$ will be close to the minimum of their respective potentials. Furthermore, because the inductive energies $E_{L_A}, E_{L_T} \gg E_J$, these minima will be close to zero and we can expand the potential $U(\mathbf{x}_A,\mathbf{x}_T)$ around $(\mathbf{x}_A,\mathbf{x}_T)=(0,0)$. Note that this expansion is used for different values of $x_{\rm ext}(t)$ for which the minimum of the $\cos()$ potential does not occur at $\mathbf{x}_A=0,\mathbf{x}_T=0$.
We discuss this approximation in more detail in \cref{sec:expansion}.
This expansion up to fourth order yields
\begin{align}
	U(\mathbf{x}_A,\mathbf{x}_T) &
\approx	\frac{E_{L_A}\mathbf{x}^2_A}{2}
	+\frac{E_{L_T}\mathbf{x}^2_T}{2}\notag\\ 
	&+E_J \cos(x_{\rm ext}(t))
	\left(
		\mathbf{x}_A \mathbf{x}_T
		-\frac{\mathbf{x}_T^2+\mathbf{x}_A^2}{2}
		+\frac{\mathbf{x}_A^2 \mathbf{x}_T^2}{4}
		-\frac{\mathbf{x}_A^3 \mathbf{x}_T+\mathbf{x}_A \mathbf{x}_T^3}{6}
		+\frac{\mathbf{x}_T^4+\mathbf{x}_A^4}{24}
	\right)\notag\\ 
	&+E_J \sin(x_{\rm ext}(t))
	\left(
		\mathbf{x}_T-\mathbf{x}_A
		+\frac{\mathbf{x}_T^2 \mathbf{x}_A -\mathbf{x}_A^2 \mathbf{x}_T}{2}
		+\frac{\mathbf{x}_A^3 - \mathbf{x}_T^3}{6}
	\right).
\end{align}
We can already see the desired coupling term, $E_J \sin(x_{\rm ext}(t)) \mathbf{x}_A^2 \mathbf{x}_T/2$.
However, there are multiple undesired additional interactions. In addition it is obvious (from the electric circuit itself) that the Hamiltonian acts the same way on target and ancilla oscillator. As will be seen in the following, a suitable choice of parameters addresses both these questions.
We first define effective, flux-dependent inductive and capacitive energies for both systems:
\begin{align*}
	&\tilde{E}_{L_m}(x_{\rm ext}(t)) = E_{L_m}-E_J\cos(x_{\rm ext}(t)), &&\tilde{E}_{C_m} = E_{C_m}\left(1-\frac{E_{C_m}}{E_{C_J}}\right) \approx E_{C_m},
\end{align*}
where the approximation comes about as $C_J \ll C_m$ for $m=A,T$.
In addition, we define (flux-dependent) frequency, and creation and annihilation operators for the two coupled oscillators:
\begin{align}
	&\omega_m(x_{\rm ext}(t)) = \sqrt{8\tilde{E}_{C_j}\tilde{E}_{L_m}(x_{\rm ext}(t))},
	&& \xi_m = {\left(\frac{2 \tilde{E}_{C_m}}{\tilde{E}_{L_j}}\right)}^{1/4}, \label{eq:freqs} \\
	&\mathbf{x}_A =	\xi_A	(a^\dag + a),
	&&\mathbf{x}_T = 	\xi_T	(b^\dag + b), \notag \\
	&\mathbf{p}_A = \i 	\frac{1}{2\xi_A}	(a^\dag - a),
	&&\mathbf{p}_T = \i 	\frac{1}{2\xi_T}	(b^\dag - b).\notag
\end{align}
All uncoupled quadratic terms in $H$ in Eq.~(\ref{eq:H}) can be put together to give a term proportional to $\omega_A  a^{\dagger} a+\omega_T b^{\dagger} b$, setting the oscillator frequencies.

In order to achieve the desired asymmetric coupling, we assume that $\xi_A \gg \xi_T$.
Because the inductance of both systems is assumed to be comparable, this implies that $\omega_A \gg \omega_T$, see \cref{tab:properties}.
In the final step, we also go to the rotating frame of both oscillators (at their frequencies $\omega_m$) and use the rotating wave approximation, \ie we only keep terms which are inherently time-independent or which are flux-dependent and oscillate with frequency $\omega_T$:
\begin{align}
	H_{\rm RWA} &\approx
	E_J \cos(x_{\rm ext}(t))
	\left(
		\frac{\xi_A^2 \xi_T^2}{2}(a^\dag a + b^\dag b + 2 a^\dag a b^\dag b)
		+\frac{\xi_A^4}{4}(a^\dag a + {(a^\dag a)}^2)
		+\frac{\xi_T^4}{4}(b^\dag b + {(b^\dag b)}^2)
	\right).\notag \\
	&+E_J \sin(x_{\rm ext}(t))
	\left(\xi_T
		\left(1-\frac{\xi_A^2}{2}-\xi_A^2 a^\dag a\right)(b^\dag \e^{\i\omega_T t} + b \e^{-\i\omega_T t})\right.\notag\\ 
		&\hspace{3cm}\left.-\frac{\xi_T^3}{6}\left(
			b^\dag\e^{\i\omega_T t}
			+b^\dag b(b^\dag\e^{\i\omega_T t}+2b\e^{-\i\omega_T t})
			+{\rm h.c.}
		\right)
	\right) \label{eq:H_RWA_orig} \\
	&\approx
	E_J \sin(x_{\rm ext}(t))
	\left(
		\xi_T \left(1-\frac{\xi_A^2}{2}\right)\left(b^\dag \e^{\i\omega_T t} + b \e^{-\i\omega_T t}\right)
		-\xi_T \xi_A^2 a^\dag a\left(b^\dag \e^{\i\omega_T t} + b \e^{-\i\omega_T t}\right)
	\right). \label{eq:H_RWA}
\end{align}
In the second approximation step, we have used $\xi_A,\xi_T \ll 1$ (dropping all fourth-order terms in $\xi_i$) and $\xi_T\ll \xi_A$ (omitting the $\xi_T^3$ term).
The $\xi_T^3$ term comes about by writing $\mathbf{x}_T^3$ in terms of annihilation and creation operators, and neglecting the parts rotating at frequency $3\omega_T$.
Although the prefactor $\xi_T^3$ is small, this term is still relevant because it will be made resonant by any drive that enables a photon-pressure coupling in the rotating frame.
In \cref{sec:nl} we will explicitly discuss the effect of the $\xi_T^3$ term.
Modulo its time-dependence, the first term of this final Hamiltonian is a known displacement that commutes with the photon-pressure coupling, the second is the traditional photon-pressure coupling Hamiltonian $\sim a^\dag a (b^\dag \e^{\i \omega t}+b \e^{-\i \omega t})$ similar to the coupling in~\cite{Johansson.etal.2014:CqedOptomechanicsTheo,Eichler.Petta.2017:CqedOptomechanicsExp}.

If the external flux is set to some constant $x_{\rm ext,0}$, only the time-independent terms remain in $H_{\rm RWA}$ and the resulting Hamiltonian is given by
\begin{align}
	H_{\rm off} \approx	E_J \cos(x_{\rm ext, 0})
	\left(
		\frac{\xi_A^2 \xi_T^2}{2}(a^\dag a + b^\dag b + 2 a^\dag a b^\dag b)
		+\frac{\xi_A^4}{4}(a^\dag a + {(a^\dag a)}^2)
		+\frac{\xi_T^4}{4}(b^\dag b + {(b^\dag b)}^2)
	\right).
\end{align}
We note that there is no photon-pressure coupling between the two modes, the only remaining non-linear terms are self-Kerr ($\sim {(a^{\dagger} a)}^2, {(b^{\dagger} b)}^2$) and cross-Kerr ($\sim a^{\dagger} a b^{\dagger} b$). The dependence of the Hamiltonian on $x_{\rm ext,0}$ means that these unwanted interactions can be turned off by setting $x_{\rm ext,0}=\pi/2$ \ie $\Phi_{\rm ext} = \Phi_0/4$.
When the photon-pressure coupling should be on and $x_{\rm ext}$ is changing over time, we do not wish to have these self-Kerr and cross-Kerr terms. We will take a flux drive so that $x_{\rm ext}(t)$ oscillates periodically around $\pi/2$ and this then directly leads to the terms proportional to $\cos(x_{\rm ext}(t))$ averaging out, see \cref{sec:drive,fig:drive}.

To turn the photon-pressure coupling on, we assume a drive such that $\sin(x_{\rm ext}(t)) = \cos(\omega_T t)$.
At first glance, such a drive seems to be difficult to achieve, as it would require a steadily increasing flux. However, one can use the symmetry of the sine around $\pi/2$ to obtain an oscillating function.
The drive is in fact a triangle wave with frequency $\omega_T/2\sim 250$MHz, an excellent approximation can easily be generated with standard equipment, see details in \cref{sec:drive}.
We insert this drive choice in \cref{eq:H_RWA} and drop all terms which remain time-dependent to obtain the desired Hamiltonian
\begin{align}
	H_{\rm on} \approx
	\frac{E_J}{2}
		\left[
			\xi_T (1-\frac{\xi_A^2}{2}) (b^\dag + b)
			-\xi_T \xi_A^2 a^\dag a(b^\dag +b)
		\right].
		\label{eq:H_eff}
\end{align}
so that the photon-pressure coupling strength $g = \frac{1}{2}E_J \xi_T \xi_A^2$.
We note that, besides the photon-pressure coupling, the Hamiltonian contains an additional displacement on the target oscillator. Since the displacement commutes with the coupling, it does not alter its effect and can be seen a systematic error on the target oscillator which can be undone by a counter-displacement.

The Hamiltonian \cref{eq:H_eff} can be easily adjusted to a photon-pressure coupling with any rotated quadrature by choosing an appropriate offset between external flux and the target oscillator.
For example, the choice $x_{\rm ext,\rm sin}(t) = x_{\rm ext}(t+\frac{\pi}{2\omega_T})$ generates a Hamiltonian of the form $H \sim \i a^\dag a (b^\dag - b)$.


The Hamiltonian $H_{\rm on}$ realizes $U_{\rm PP} = S_{q_T}^{a^\dag a}$ (modulo the unconditional displacement), where the photon number operator only has non-negative eigenvalues.
Therefore, if we view this interaction as an ancilla-oscillator dependent displacement on the target oscillator, all displacements $S_{q_T}^{a^\dag a}$ point in the same direction, and the post-measurement state in the target oscillator will be off-center in phase space and contain an unnecessarily high number of photons.

In order to reduce the photon number, one can apply a displacement drive such that the unconditional displacement during the interaction is $S_{q_T}^{-\braket{a^\dag a}/2} = Z^{-\braket{a^\dag a}}$.
The idea is the same as for phase estimation when using qubits as ancilla, see~\cite{TW:GKP}. We will use such a counter-displacement in all numerical simulations in this paper.

One thing to observe is that the frequency of the ancilla (and the target) oscillator depends on the flux drive through Eq.~(\ref{eq:freqs}), hence we are working in a flux-dependent rotating frame which has to be carefully tracked (in order to read out the phase of the ancilla oscillator and do additional counter-displacements on the ancilla oscillator).

In some settings, it might be desirable to use a drive $\sin(x_{\rm ext})(t) = 1-\delta+\delta\cos(\omega_T t), 0<\delta\leq 1$.
It is possible to do so, and a drive with $\delta<1$ is easier to generate, but this costs some coupling strength, see \cref{sec:drive} for details.
In the main text, we will use the maximal possible coupling strength \ie $\delta=1$ unless mentioned otherwise.

The values for resonance frequency, coupling strength and the leading order error terms for a typical setup are given in \cref{tab:properties}.
In order to maximize the coupling strength, it is beneficial to reduce the Josephson energy while simultaneously increasing the inductances of both circuits in order to keep $E_J\ll E_{L_m}$.
Furthermore, it is beneficial to make the inductance of the target oscillator smaller than that of the ancilla oscillator:
The ratio between the third order nonlinearity and the photon-pressure coupling strength is proportional to the ratio of the inductances.
For a Josephson energy around $10$ GHz and an inductance of the ancilla oscillator around $2$ nH, a coupling strength $g/(2\pi)$ well above $10$ Mhz can be achieved.
Note that the Kerr and cross-Kerr effects on both oscillators might be large during the interaction due the first term in Eq.~(\ref{eq:H_RWA}), however they both oscillate in sign and will therefore be echoed out (see \cref{sec:drive}).

\subsection{Release of Ancilla Oscillator State}
\label{sec:release}
In order to meet both the demands of fast read-out and low photon loss, it is desirable to be able to effectively turn the ancilla oscillator decay rate from low to high.
There are a few ways to achieve this, for example with a tunable inductive coupling~\cite{yin:catch-release}, a frequency tunable oscillator~\cite{pierre+:storage} a pump-tunable beam splitter to a lossy oscillator~\cite{pfaff+:controlled-release} or a parametric coupler~\cite{flurin+:node}.
Note that most of these references work towards catch and release schemes, hence if the tunable coupling is simply used for readout the achieved fidelities can be expected to be larger. In particular, the Q-switch scheme in~\cite{pfaff+:controlled-release} in which a pump is used to temporally frequency-match the ancilla oscillator with a lossy oscillator seems attractive.
In this work, the ratio between the closed and open decay rates is about 1000: the authors increase the effective life-time of an oscillator from about $0.5$ms to $0.5\mu$s, with efficiency exceeding 98\%. The paper reports that the coherence and phase of oscillator states with up to 5 photons can be well resolved.

In the protocol presented here, it is also possible to use the fact that the ancilla oscillator has a tunable frequency. If a lossy fixed-frequency oscillator is placed between transmission line and ancilla qubit, the ancilla can be brought into resonance with it, increasing the decay rate. Note that this idea is as in~\cite{pierre+:storage}, but reversing the roles of the frequency-tunable and fixed-frequency oscillator.
The lossy oscillator needs to be off-resonance, effectively acting as Purcell filter, except during readout. An advantage of this approach is that it does not require any further circuit elements.
As an example, consider an ancilla oscillator with properties as in~\cref{tab:properties}.
In this case, the resonance frequency is between $f(x_{\rm ext}=0) = 9.75$ GHz and $f(x_{\rm ext}=\pi) = 10.25$ GHz. If the lossy resonator has resonance frequency $9.75$ GHz and we want lossy oscillator and ancilla oscillator to be separated by at least $250$ MHz, we require that $\pi/2\leq x_{\rm ext}\leq \pi$. This can be achieved by modifying the drive during the interaction, see \cref{sec:drive}. After the interaction time, we set the flux to $x_{\rm ext}=0$ in order to bring the ancilla oscillator into resonance with the lossy oscillator.


\section{Modeling The Modular Quadrature Measurement}
\label{sec:meas}
In this section we derive the effective squeezing after the protocol, averaged over all possible measurement outcomes, as a function of the number of photons in the ancilla oscillator.
Our measurement model could be made more precise by including a description of the release mechanism discussed in~\cref{sec:release}, but this does not change the main idea as long as the coherent state is heterodyne-measured at the end.
In \cref{app:time-int} we look at another aspect of the actual measurement as it is performed in the circuit-QED lab, namely the measurement outcome is only obtained as a time-integrated process on outgoing radiation which is leaking out of the lossy oscillator (which is in turn coupled to oscillator $A$ via the switch discussed in \cref{sec:release}). We verify that using the correct time-integration filter leads to no additional noise resulting in the same effective squeezing due to the measurement.

\subsection{Effective Squeezing}
\label{sec:toy}
We will analyze a measurement of the $S_q$ stabilizer using the photon-pressure interaction $U_{\rm PP}$ in \cref{eq:U_{TA}}. A similar measurement of $S_p$ will commute with the measurement of $S_q$ and will have identical features. We drop the label $T$ from $\hat{q}_T$ acting on the target oscillator from now on.

After the photon-pressure interaction with the target oscillator the goal is to measure the Husimi Q-function $Q(\beta)=\frac{1}{\pi} \bra{\beta} \rho \ket{\beta}$ of the ancilla oscillator in single-shot fashion~\cite{eichler:detection}.
Such a ``heterodyne'' measurement of an oscillator can be modeled as a projective measurement in the overcomplete basis of coherent states~\cite[p.24]{book:Wiseman.Milburn:measurement}.
The resulting coherent amplitude $\beta$ has a real $\Re(\beta)$ ($\propto$ ``I'') and imaginary part $\Im(\beta)$ ($\propto$ ``Q'') and will leave some target oscillator state $\rho_{\beta}$. Using this measurement outcome $\beta=|\beta|\exp(\i\varphi)$, one infers that the eigenvalue of $S_q$ is $\exp(\i \varphi)$.
The uncertainty in this phase is captured by the phase variance which relates directly to the effective squeezing of $S_q$.\\

We assume that the initial state of the ancilla oscillator is a coherent state $\ket{\alpha}$ with $\alpha \in \mathbb{R}$. If we would apply a heterodyne measurement directly to a coherent state $\ket{\alpha}$, we expect that its outcome $\beta\in\mathds{C}$ will be concentrated around $\alpha$.
In our scenario, when we apply such measurement after the interaction $U_{\rm PP}$, we obtain a measurement operator $M_{\beta} \equiv M_{\beta}(\alpha)$ corresponding to measurement result $\beta$ as
\begin{align}
	M_{\beta}(\alpha)  &= \frac{1}{\sqrt{\pi}}\bra{\beta}_A U_{\rm PP}\ket{\alpha}_A\\
	\int_{\mathds{C}} \mathrm{d}^2 \beta\; M_\beta^\dag M_\beta &=  
	\frac{1}{\pi} \int_{\mathds{C}} \mathrm{d}^2 \beta\; \notag \bra{\alpha}_A U_{\rm PP}^\dag\ket{\beta}_A\bra{\beta}_A U_{\rm PP}\ket{\alpha}_A \\ 
	&= \bra{\alpha}_A U_{\rm PP}^\dag U_{\rm PP} \ket{\alpha}_A = \bra{\alpha}_A \mathds{1}_{TA} \ket{\alpha}_A = \mathds{1}_T, \notag
\end{align}
We can evaluate the measurement operator explicitly, using that $\bra{\beta} \alpha\rangle=\exp(-\frac{1}{2}|\alpha-\beta|^2)\exp(\frac{1}{2}(\beta^* \alpha-\beta \alpha^*))$, giving
\begin{equation}
M_{\beta}=\frac{1}{\sqrt{\pi}}\bra{\beta} \alpha e^{\i 2 \sqrt{\pi}\hat{q}_T}\rangle=\frac{1}{\sqrt{\pi}}
\exp(-\frac{1}{2}|\alpha e^{\i2\sqrt{\pi}\hat{q}}-\beta|^2)\exp\left(\frac{1}{2}(\alpha(\beta^* e^{i2 \sqrt{\pi} \hat{q}}-\beta e^{-\i 2 \sqrt{\pi} \hat{q}}))\right).
\end{equation}
When we apply this to an initial input state $\rho_{\rm in}$ in the target oscillator, the output state will be $\rho_{\beta}=\frac{1}{\mathbb{P}(\beta)} M_{\beta} \rho_{\rm in}M_{\beta}^{\dagger}$.
The probability for outcome $\beta$ with initial state $\rho_{\rm in}=\iint_{\mathbb{R}^2} \diffd q\ \diffd q' \,\rho_{\rm in}(q,q') \ket{q}\!\bra{q'}$ as input is given by
\begin{equation}
\mathbb{P}_{\rho_{\rm in}}(\beta)= \Tr (M_{\beta}^{\dagger} M_{\beta} \rho_{\rm in} )=
\int_{\mathbb{R}} \diffd q \; \rho_{\rm in}(q,q) \exp(-|\alpha e^{\i 2\sqrt{\pi}q}-\beta|^2),
\label{eq:conc}
\end{equation}
showing that $\beta$ is concentrated around the rotated $\alpha$. Figure~\ref{fig:wigner-simple} shows this probability $\mathbb{P}_{\rm vac}(\beta)$, starting with $\bar{n}=|\alpha|^2=3$ and $\rho_{\rm in}$ the vacuum state. It also shows the Wigner function of the resulting state $\rho_{\beta}$ for which $\mathbb{P}_{\rm vac}(\beta)$ is maximal.
Using the definition \(\varphi\equiv\arg(\beta)\), an alternative way of writing $M_{\beta}$ is
\begin{align}
	M_{\beta}
	&= \frac{1}{\sqrt{\pi}}\exp(-\frac{1}{2}(|\alpha|^2+|\beta|^2))\exp(K_{|\beta|}\cos(2\sqrt{\pi}\hat{q}-\varphi)/2)\exp(\i K_{|\beta|}\sin(2\sqrt{\pi}\hat{q}-\varphi)/2), \label{eq:M_direct}
\end{align}
defining the concentration parameter
\begin{equation}
 K_{|\beta|}=2|\alpha \beta|.
\end{equation}
This leads to
\begin{equation}
	M_\beta^\dag M_\beta = \frac{1}{\pi}\exp(-|\alpha|^2-|\beta|^2)\exp(K_{|\beta|}\cos(2\sqrt{\pi}\hat{q}-\varphi)).
\label{eq:M}
\end{equation}

\begin{figure}[htb]
	\includegraphics[width=0.4\textwidth]{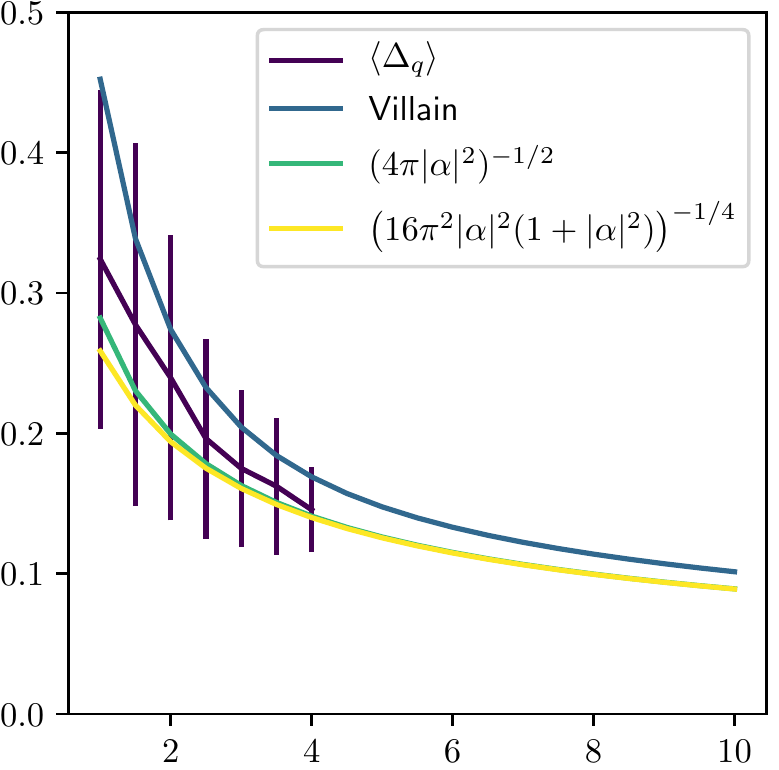}
	\caption{Purple: We numerically simulate the average amount of squeezing $\Delta_q$ (see \cref{def:squeezing}) obtained using a coherent state $\ket{\alpha}$ with $\bar{n}=|\alpha|^2$ photons to measure $S_q$ on a vacuum input state. In more detail, we generate a $\beta$ and $\rho_{\beta}$ and calculate $\Delta_q(\rho_{\beta})$, the error bars indicate the standard deviation over different measurement results $\beta$.
	Blue: Mean effective squeezing estimate according to \cref{eq:Villain}, using the Villain approximation to evaluate the expectation value for the sharpness on a vacuum state.
	Green: A simple approximate expression for the mean effective squeezing is $\braket{\Delta_q}\approx 1/\sqrt{4 \pi |\alpha|^2}$.
	Yellow:  A lower bound on the green curve which replaces the average value $\langle |\beta| \rangle$ by $\sqrt{\langle |\beta|^2 \rangle}$.
	Overall, the mean squeezing parameter goes down as $1/\sqrt{\bar{n}}$ where $\bar{n}$ is the average number of photons in the ancilla state used to implement the modular $q$-measurement.}
\label{fig:sq-result}
\end{figure}

Because the measurement outcome is random, we are interested in the mean effective squeezing of the final state $\rho_{\beta}$, averaged over all possible outcomes $\beta$. This is hard to compute in the general case, although it can easily be evaluated numerically, see Fig.~\ref{fig:sq-result}.
Analytically, even for a vacuum state input, the computation of the mean effective squeezing $\langle \Delta_q \rangle=\int d\beta \,\Delta_q (\rho_{\beta})$ is non-trivial.
For this reason, we consider the mean or average \emph{sharpness} which equals $|{\rm Tr} S_q\rho_{\beta}|$ averaged over different outcomes $\beta$, that is, we focus on estimating
\begin{equation}
\langle |{\rm Tr} S_q| \rangle \equiv \int \diffd\beta \;\mathbb{P}(\beta)|{\rm Tr} S_q \rho_{\beta}|.
\end{equation}
It should be observed that $\int \diffd\beta\; |{\rm Tr} S_q \rho_{\beta}| \neq |\int \diffd\beta\; {\rm Tr} S_q \rho_{\beta}|=|{\rm Tr} S_q \rho_{\rm in}|$ as ${\rm Tr} S_q \rho_{\beta}$ is complex. 
Since $S_q$ commutes with $M_{\beta}$ we have
\begin{align}
\langle |{\rm Tr} S_q| \rangle & =  \int_{\mathbb{C}} \diffd\beta\; | \bra{\rm vac} S_q M_{\beta}^{\dagger} M_{\beta} \ket{\rm vac}|\nonumber \\  
& = \frac{1}{\pi \sqrt{\pi}} \int \diffd\beta\; \exp(-|\alpha|^2-|\beta|^2) \left|\int \diffd q\; \exp(-q^2) \exp(\i 2\sqrt{\pi}q)  
\exp(K_{|\rm \beta|}\cos(2\sqrt{\pi}q-\varphi))\right|.  
\label{eq:Sq_vac}
\end{align}
At $b \geq 2$, one can use the convenient Villain approximation $\exp(b \cos(x)) \approx \sum_{n \in \mathbb{Z}} \exp(b) \exp(-\frac{b}{2}{(x-2\pi n)}^2)$~\cite{JK:villain}.  
For large $K_{|\beta|}$, the dominant contribution comes from small values of $|n|$: for $K_{|\beta|} \geq 2$ one can restrict the sum to $n=0,\pm 1,\pm 2$ with $-\pi \leq \varphi \leq \pi$.

If we assume that the outcomes of $\beta$ are concentrated around values where the Villain approximation holds (which is reasonable since we know that $\mathbb{P}(\beta)$ is concentrated around $|\beta|=\alpha$ from Eq.~(\ref{eq:conc})), then one can apply this approximation and evaluate the resulting Gaussian integral to get
\begin{align}
\langle |{\rm Tr} S_q| \rangle \approx \frac{1}{\pi\sqrt{\pi}} \int_{|\beta|_c}^{\infty} \diffd|\beta|\; |\beta| \exp(-{(|\alpha|-|\beta|)}^2)\frac{\exp(-\pi)}{\sqrt{2 K_{|\beta|}}} \int_{-\pi}^{\pi} \diffd\varphi\; \left|\vartheta_3(\i \pi-\varphi/2,\exp(-\pi-1/(2K_{|\beta|})))\right|. 
\label{eq:Villain}
\end{align}
Here $\vartheta_3(z,q)=\sum_{n\in \mathbb{Z}} q^{n^2} e^{2 \i n z }$ is the theta function and $|\beta|_c$ is a lower cut-off to allow for the Villain approximation.
The lower cut-off $|\beta|_c$ is chosen such that firstly $\mathbb{P}(|\beta| < |\beta|_c)\ll 1$, and secondly $|\alpha \beta|_c \geq 1$ to allow for the Villain approximation with $K_{|\beta|}\geq 2$.
We take $|\beta|_c=1/|\alpha|$ so that for $\bar{n}\geq 5$ the probability for such $|\beta|_c$ is low (suppressed by $\exp(-{(5-1/\sqrt{5})}^2 \approx 0.04$).
The function $\vartheta_3(z,q)$ is oscillatory with $n$, but contributions beyond $n=0,\pm 1,\pm 2$ are negligible. Inserting the mean sharpness with its approximation in \cref{eq:Villain} in the expression for $\Delta_q$, we obtain the red curve in \cref{fig:sq-result}.

We can also consider the eigenvalue phase of $S_q$ of the output state $\rho_{\beta}$, i.e.
\begin{equation}
\arg({\rm Tr} S_q \rho_{\beta})=\arg\left(\int \diffd q\; \rho_{\rm in}(q,q) S_q \exp(K_{|\beta|}\cos(2\sqrt{\pi}q-\varphi))\right).
\label{eq:estimate}
\end{equation}
When $\rho_{\rm in}(q,q)$ is a uniform distribution, \ie there is no prior bias for $\hat{q}$ (or $S_q$), the integral over $q$ results in $\exp(i \varphi)$, confirming that the best choice for inferring the eigenvalue of $S_q$ is indeed $\exp(\i \varphi)$.
If we have prior information on the input state to the measurement, \eg it is the vacuum state, then one can use \cref{eq:estimate} as the outcome of the measurement.

As a sanity check we examine $\langle |{\rm Tr} S_p | \rangle$ (or $\langle |{\rm Tr} X | \rangle$) after the modular $q$-measurement. First of all, note that the expectation $\int \diffd \beta\; {\rm Tr} S_p \rho_{\beta}$ is unchanged, since $S_p$ commutes with the $S_q$-measurement, so $\int \diffd\beta\; \Tr S_p \rho_{\beta}=\Tr S_p \rho_{\rm in}$. 
Thus for the output state, the squeezing of $S_p$ (or $X$) is unchanged as expected. In addition, if we consider the mean sharpness we can also see it is preserved when we start with the vacuum state:
\begin{equation}
\langle |{\rm Tr} S_p | \rangle=\int \diffd\beta\; |\bra{\rm vac} S_p M_{\beta}^{\dagger} M_{\beta}  \ket{\rm vac}|=\int\diffd \beta\; \bra{\rm vac} S_p M_{\beta}^{\dagger} M_{\beta}  \ket{\rm vac}=\bra{\rm vac} S_p \ket{\rm vac}=\exp(-\pi).  
\end{equation}
The second equality in the equation above follows from the fact that $\bra{\psi} S_p M_{\beta}^{\dagger} M_{\beta} \ket{\psi}$ for any state $\ket{\psi}$ whose wave function is nonnegative in the $q$-basis, i.e. $\psi(q) \geq 0$, so we can omit the absolute value and use $\int \diffd\beta\; M_{\beta}^{\dagger} M_{\beta}=I$. 
One should observe that the preservation of the mean sharpness does not automatically follow from the commutation of $S_p$ with $M_{\beta}$ or $M_{\beta}^{\dagger}$.

\subsection{Measurement Squeezing Strength}
If the initial state in the target oscillator is arbitrary, it is not possible to analytically evaluate the mean sharpness in \cref{eq:Sq_vac}.
Moreover, we are interested in a quality measure of the measurement protocol which is \emph{independent} of the initial state. To address this, we can use that the parameter $K_{|\beta|}$ has a very simple relation to the effective squeezing.
If we assume a uniform distribution as initial state, then the final state of the protocol will be of the form $\ket{\psi_\beta}\propto \int \diffd q\; M_\beta \ket{q}$.
Using Eq.~(\ref{eq:M_direct}) we see that the outgoing wave function has probability distribution $\mathbb{P}(q) \propto M_{\beta}^{\dagger} M_{\beta}$, proportional to a von-Mises probability density $P_{VM}(x)$ with angle variable $x=2\sqrt{\pi} q \mod 2\pi$ with mean $\varphi$ and concentration $K_{|\beta|}$. The variance of the von-Mises distribution is given by $1/K_{|\beta|}$ for large $K_{|\beta|}$.
If we convert this to an effective squeezing in $q$, it gives $\Delta_q \approx \sqrt{1 / (2 \pi K_{|\beta|})}$.

Thus, computing the expected value for $|\beta|$, giving $\langle K_{|\beta|}\rangle=2 \alpha \langle |\beta| \rangle$, gives a measure of how effectively squeezed the outgoing state will be.
Not surprisingly, one can show, see the mathematical details in the following paragraph, that $\langle |\beta| \rangle \approx |\alpha|$ so that $\Delta_q \approx 1/\sqrt{4\pi |\alpha|^2}$.
Since $\langle |\beta| \rangle \leq \sqrt{\langle |\beta|^2 \rangle}$, we can also use a squeezing lower bound which reads $\sqrt{1/(4\pi |\alpha| \sqrt{1+|\alpha|^2})}$ using that $\langle |\beta|^2 \rangle=1+|\alpha|^2$ (see below). Fig.~\ref{fig:sq-result} shows that these state-independent bounds are in good agreement with numerics as well as our analytical approximation when the input state is the vacuum state.\\

To estimate $\langle f(|\beta|) \rangle$ where $f(x)$ is some function, we note the following useful property which we prove as a lemma:
\begin{lem}
The input state in the target cavity $\rho_{\rm in}$ does not influence the expectation of any function $f(|\beta|)$ where $\beta$ is the outcome of measuring in the ancilla mode in an overcomplete coherent basis.

\label{lem:indep}
\begin{proof}
For a general input state $\rho_{\rm in}$ we have
\begin{equation}
\langle f(|\beta|)\rangle=  \int \diffd\beta\; \mathbb{P}_{\rm in}(\beta) f(|\beta|)=\int_0^\infty \diffd|\beta|\; |\beta| f(|\beta|) \int_{-\pi}^{\pi} \diffd\varphi\;\mathbb{P}_{\rm in}(|\beta|e^{i\varphi})=  
\int_0^\infty \diffd|\beta|\; |\beta| f(|
\beta|) \int_{-\pi}^{\pi} \diffd\varphi \;{\rm Tr} M_{\beta}^{\dagger} M_{\beta} \rho_{\rm in}. 
\end{equation}
We can use the Jacobi-Anger expansion
\begin{equation}
\exp(i z \cos(\theta))=\sum_{n\in \mathbb{Z}} \i^n J_n(z) e^{\i n \theta}=J_0(z)+2 \sum_{n=1}^{\infty} \i^n J_n(z) \cos(n\theta),
\end{equation}
where $J_n(z)$ is the Bessel function of the first kind and using $J_{-n}(z)={(-1)}^n J_n(z)$. The modified Bessel function of the first kind $I_n(z)$ is defined as $I_n(z)=\i^{-n}J_n(\i z)$ and it follows that
 $\exp(b \cos(x))=\sum_{n\in \mathbb{Z}} I_n(b)\exp(\i n x)$ where $I_n(b)$ is the modified Bessel function of the first kind of order $n$.
We can then use Eq.~(\ref{eq:M}) to write
\begin{equation}
 \int_{-\pi}^{\pi} \diffd\varphi\; {\rm Tr} M_{\beta}^{\dagger} M_{\beta} \rho_{\rm in}=\frac{1}{\pi} 
\exp(-|\alpha|^2-|\beta|^2)\sum_{n \in \mathbb{Z}} \int \diffd q\; \rho_{\rm in}(q)I_n(K_{\beta})\exp(\i n 2 \sqrt{\pi} q)
\int_{-\pi}^{\pi} \diffd\varphi\; \exp(-\i n \varphi). 
\label{eq:cancel}
\end{equation}
The integral over $\varphi$ leads to $n=0$ being the only surviving term in $\sum_{n \in \mathbb{Z}}$, thus removing all dependence on $\rho_{\rm in}$ in the integral over $q$. Hence
\begin{equation}
\langle f(|\beta|)\rangle=  2\exp(-|\alpha|^2) \int_0^\infty  \diffd|\beta|\; |\beta| f(|
\beta|)\exp(-|\beta|^2)I_0(K_{\beta}),
\label{eq:gen-exp}
\end{equation}
independent of $\rho_{\rm in}$.
\end{proof}
\end{lem}

Equation (\ref{eq:gen-exp}) allows us to get an expression for $\langle |\beta| \rangle$ as
\begin{align}
\langle |\beta| \rangle=
2\exp(-|\alpha|^2) \int_0^\infty \diffd |\beta|\;  |\beta|^2  \exp(-|\beta|^2)I_0(2\alpha|\beta|)=
\frac{\sqrt{\pi}}{2}\e^{-|\alpha|^2/2}\left(I_0(|\alpha|^2/2) + \alpha^2\left(I_0(|\alpha|^2/2)+I_1(|\alpha|^2/2)\right)\right),
\label{eq:beta_av}
\end{align}
which for $\alpha \geq \sqrt{2}$ is virtually indistinguishable from $\langle |\beta| \rangle \approx \alpha$, as expected.
Therefore, the expected effective squeezing can be approximated as $\langle \Delta_q \rangle \approx 1/\sqrt{4\pi |\alpha|^2}$ as plotted in Fig.~\ref{fig:sq-result}. Fluctuations around this expected value are determined by
\begin{align}
\langle |\beta|^2 \rangle=2\exp(-|\alpha|^2) \int_0^\infty \diffd |\beta|\; |\beta|^3  \exp(-|\beta|^2)I_0(2\alpha|\beta|) =1+|\alpha|^2.
\label{eq:fluc}
\end{align}
so that ${\rm Var}(|\beta|)=\langle {(|\beta|-\langle |\beta| \rangle)}^2 \rangle \approx 1$.

As expected, these statistics are identical to that of a direct overcomplete measurement in the coherent basis applied to a state $\ket{\alpha}$, \ie without any coupling to a target oscillator. The only dependence on $\rho_{\rm in}$ is found in the phase $\varphi$.
In conclusion, the amplitude of the measurement result $|\beta|$ correlates with the accuracy of the measurement, the phase gets more precisely resolved the larger the measured coherent state is.
Thus, the expectation value $\braket{|\beta|}$ gives an indirect, but easily accessible way to estimate the effective squeezing by the measurement.

\section{Noise and Imperfections}
\label{sec:noise}

As compared to a perfect heterodyne measurement of the rotated coherent state in the ancilla oscillator, there will be several sources of loss and imperfections in the modular quadrature measurement.
In the sections below, we discuss the effect of photon loss on the ancilla and target oscillators as a change in the effective squeezing parameters.
Importantly, photon loss on the ancilla oscillator during the photon-pressure coupling is an immediate cause for feedback dephasing errors, similar as when preparing a grid state via coupling to a transmon ancilla qubit~\cite{TW:GKP}. Loss in read-out in the heterodyne measurement state simply reduces the effective $\alpha$ that is used in the protocol, diminishing the strength of the measurement.

After the discussions on photon loss, we investigate the leading nonlinear term acting on the target oscillator in \cref{sec:nl}. As the nonlinear term only acts during the interaction of the target and ancilla oscillators, it acts as an additional unitary operation. We discuss and numerically simulate its effect as a change of the effective squeezing parameters.

Finally, we investigate the effect of flux noise during the interaction, as the coupling Hamiltonian between the target and ancilla oscillators depends on an external flux.
A small, quasi-static flux offset has the effect that the measured quadrature is slightly rotated, \ie a flux offset $\epsilon$ means that the photon-pressure Hamiltonian is changed to $\tilde{H}_{\rm PP} \sim a^\dag a(\cos(\epsilon)\hat{q} \pm \i\sin(\epsilon)\hat{p})$.
We will see that the parametric drive already provides a first order correction to this type of noise because the sign in the modified Hamiltonian $\tilde{H}_{\rm PP}$ changes with frequency $\omega_T$, which is large compared to $1/t_{\rm coupl}$.

\subsection{Photon Loss in Ancilla Oscillator during Photon Pressure Interaction }
\label{sec:loss-ancilla-int}

Imagine that prior to the heterodyne measurement to measure $S_q$, but during the action of the photon-pressure coupling $U_{\rm PP}$, photon loss occurs from the ancilla resonator at rate $\kappa_c$.
This error will feedback to the target oscillator as a dephasing error in the $\ket{q}$ basis and such a dephasing error will affect $\Delta_p$. In addition, photon loss affects the quality of the $S_q$ measurement itself by effectively reducing the amplitude of the coherent state which is used in the measurement.

We assume that we are in the targeted regime, in which there is at most a single photon loss error in a time $t_{\rm coupl}$, or $\kappa_c t_{\rm coupl} \alpha^2 \ll 1$. Let $\gamma=\kappa_c t_{\rm coupl}$.
The no-photon loss operator $E_0=I-\gamma \hat{n}/2 \approx \exp(-\gamma \hat{n}/2)$ commutes with the evolution of $H_{\rm PP}$, but the single-photon loss operator $E_1=\sqrt{\gamma} \,a$ does not. Hence the state of ancilla and target oscillator at time $t$ is
\begin{align}
\rho(t)&=	\exp(-\gamma\hat{n}/2-\i 2 \sqrt{\pi}\hat{q} \hat{n})\rho_{\rm in} \otimes \ket{\alpha}\!\bra{\alpha} \exp(-\gamma\hat{n}/2+\i 2 \sqrt{\pi}\hat{q} \hat{n}) \nonumber \\
&\hspace{1cm}+\kappa_c\int_{0}^{t_{\rm coupl}} \diffd t\; A(t) \rho_{\rm in} \otimes \ket{\alpha}\!\bra{\alpha} A^{\dagger}(t), \nonumber \\
A(t)&=\exp(-i (2 \sqrt{\pi}-t \sqrt{2} g) \hat{q} \hat{n}) a \exp(-i t \sqrt{2} g \,\hat{q} \hat{n}).
\label{eq:photon_loss}
\end{align}
When we apply the heterodyne measurement to the ancilla oscillator and obtain outcome $\beta$, we thus apply to $\rho_{\rm in}$ the transformation
\begin{align*}
\rho_{\rm in} \rightarrow \rho_{\beta}&=(1-\alpha^2\gamma) M_{\beta}(\alpha \exp(-\gamma)) \rho_{\rm in} M_{\beta}^{\dagger}(\alpha \exp(-\gamma)) \nonumber \\
&\hspace{0.5cm}+\alpha^2 \gamma M_{\beta}(\alpha) \left[\frac{1}{t_{\rm coupl}}\int_0^{t_{\rm coupl}} \diffd t\; \exp(-\i \sqrt{2} g \hat{q} t) \rho_{\rm in} \exp(\i \sqrt{2} g \hat{q} t)\right] M^{\dagger}_{\beta}(\alpha).
\end{align*}
The last term can be viewed as applying, with probability $\sim \alpha^2 \gamma$, a mixture of shift errors with an average shift of strength $\sqrt{2} g t_{\rm coupl}/2=\sqrt{\pi}$.
This dephasing feedback error tends to localize the $q$-quadrature, hence affecting the extent to which the state can be an eigenstate of $S_p$ or $X$. The average feedback shift error upon photon loss is a logical shift $Z$, immediately leading to the loss of the logical information.
We can explicitly look at the effect of such photon loss when $\rho_{\rm in}=\ket{\rm vac}\bra{\rm vac}$.
Since the expression for ${\rm Tr} S_q \rho_{\beta}$ for any input state $\rho_{\rm in}$ only involves diagonal terms $\ket{q}\bra{q}$, the dephasing in the $q$-basis due to photon loss has no effect.
This means that we can view such loss as occurring after the interaction, simply leading to $\ket{\alpha} \rightarrow \ket{\alpha \exp(-\gamma/2)}$.
This loss affects the measurement quality in the same way as any readout loss, see \cref{sec:read-out}.
We can consider the effect of the feedback error on $\Delta_p$ as follows.
After the $S_q$ measurement with outcome $\beta$ we consider the expected eigenvalue sharpness of $S_p$ (or, similarly $X$)
For this we need to evaluate
\begin{align}
\int \diffd\beta\; |{\rm Tr} S_p \rho_{\beta}|&=\int \diffd\beta\; \left|(1-\alpha^2\gamma) {\rm Tr}S_p  
M_{\beta}^{\dagger}(\alpha \exp(-\gamma)) M_{\beta}^{\dagger}(\alpha \exp(-\gamma))\rho_{\rm in}\right. \nonumber \\ 
&\hspace{1cm}+\left. \alpha^2 \gamma{\rm Tr} \left[\frac{1}{t_{\rm coupl}} \int \diffd t\; \exp(\i \sqrt{2} g \hat{q} t) S_p \exp(-\i \sqrt{2} g \hat{q} t)\right]
M_{\beta}^{\dagger}(\alpha)M_{\beta}(\alpha) \rho_{\rm in}\right|.
\label{eq:exp-Sp}
\end{align}
The commutation relation $\exp(\i u \hat{q}) \exp(-\i 2 \sqrt{\pi} \hat{p})=\exp(-\i 2 \sqrt{\pi} \hat{p}) \exp(i u \hat{q}) \exp(\i 2 \sqrt{\pi} u)$ can be used to do the averaging integral over $t$ which leads to the contribution from the single-photon loss term to be zero.
This essentially means that upon the loss of an actual photon the eigenvalue phase of $S_p$ is fully randomized. The expected value for $X$, \ie $\int d\beta |X \rho_{\beta}|$ suffers similarly, \ie upon the actual loss of a photon the eigenvalue of $X$ gets fully randomized. The randomization leads to
\begin{equation}
\int \diffd\beta\; |{\rm Tr} S_p \rho_{\beta}|=(1-\alpha^2\gamma) \int \diffd\beta\; \left| {\rm Tr}S_p  
M_{\beta}^{\dagger}(\alpha \exp(-\gamma)) M_{\beta}^{\dagger}(\alpha \exp(-\gamma))\rho_{\rm in}\right|=(1-\alpha^2\gamma) |{\rm Tr} S_p \rho_{\rm in}|,
\end{equation}
where the last equality follows immediately when the wavefunction of $\rho_{\rm in}$ is real in the $q$-basis (as is the case for a vacuum state). One can also observe that $|\int d\beta S_p \rho_{\beta}|=(1-\alpha^2\gamma) |{\rm Tr} S_p \rho_{\rm in}$, since due to the photon loss $S_p$ no longer commutes with the $S_q$ measurement.

As conclusion, we have the following. Imagine that we started the modular measurement of $q$ with a state with squeezing parameter $\Delta_p < 1$. After the measurement we obtain an enhanced $\tilde{\Delta}_p \approx \sqrt{\frac{\alpha^2 \gamma}{\pi}+\Delta_p^2}$, showing how the feedback error negatively affects the squeezing in $p$.

\subsection{Comparison with Sequential-Qubit Phase Estimation Measurement and Photon Loss on Target Oscillator}
\label{sec:compare}

Previous work has analyzed how to measure the eigenvalue of $S_q$ (or $S_p$) via coupling the target oscillator with a sequence of qubits, using a qubit-controlled displacement interaction, followed by qubit measurement.
In this scheme, each qubit measurement (via a read-out oscillator) provides at most 1 bit of information.
For this sequential qubit read-out, one can use a tunable longitudinal interaction between transmon qubit and storage cavity of the form $\sqrt{2} g \frac{I-Z}{2} \hat{q}$.
This form of the coupling implies that the interaction time $t_{\rm coupl}$ is the same value as in the photon-pressure protocol with a large coherent state. If the ancilla oscillator is harmonic, one can use the vacuum state $\ket{0}$ and Fock state $\ket{1}$ as the two qubit states.
Hence, the longitudinal interaction is merely the photon-pressure coupling applied to these Fock states. However, the input state of this sequential scheme and the subsequent measurement of the qubits cannot be directly mapped onto the photon-pressure scheme using a coherent state.

To compare the sequential qubit scheme with the proposed modular quadrature measurement, we have to separately discuss the two dominant sources of error, photon loss on the ancilla oscillator and photon loss on the target oscillator.
With respect to photon loss on the ancilla oscillator: an important possible advantage of the photon-pressure scheme proposed is that a single oscillator-measurement is used instead of a sequence of qubit measurements, making it possible that the photon-pressure scheme is much faster.
This would lead to lower photon loss error rate on the target oscillator (as it is waiting while the ancillary system is being measured).
To compare times, in~\cite{pfaff+:controlled-release} the release and measurement take time $O(1) \mu$s while in the same set-up the high-fidelity single transmon qubit measurement took a similar amount of time.
If we use a coherent state with $\bar{n}=3$, \cref{fig:sq-result} shows that one can obtain $\Delta_q \approx 0.18$ assuming no losses. Data from~\cite{DTW:sensor} show that one needs at least $M=12$ rounds to get to $\Delta_q=0.2$. Also, in~\cite{CI+:grid} a grid state was stabilized after about 20 rounds of qubit measurements of duration 600 ns (including losses) to $\sigma=0.16$ which corresponds to $\Delta=0.22$ here.

With respect to photon loss on the ancilla qubit or oscillator, one can make the following observations. First, note that in the sequential execution of a protocol using ancilla qubits, arguments can be made that the squeezing parameter $\Delta_q$ will decrease as $1/\sqrt{M}$ where $M$ is the number of rounds in phase estimation protocol~\cite{TW:GKP, DTW:sensor}.
Then, similar as in the photon-pressure protocol, for each qubit measurement, there is a probability $\gamma=\kappa_c t_{\rm coupl}$ for amplitude damping (\ie photon loss) and hence a feedback error which fully randomizes the eigenvalue of $S_p$ or $X$.
Hence after $M$ such rounds the probability for a $Z$ error scales as $\sim \gamma M \sim \gamma/\Delta^2$. In our proposed strong measurement scheme, the error probability is $\gamma \bar{n} \sim \gamma/ \Delta^2$, showing that both schemes effectively have the \emph{same} tradeoff in having a higher logical $Z$ error probability when targeting a smaller $\Delta$.
It is thus a matter for what $\bar{n}$ one has $\kappa_c t_{\rm coupl} \bar{n} \ll 1$ which determines whether a strong measurement with $\bar{n} > 1$ is most effective.

In this context, it should also be noted that it is not the aim for a GKP state preparation protocol to necessarily prepare the highest possible $\Delta$.
Photon loss on the target oscillator during the protocol and during measurement of the ancillary system will lead to drift and diffusion of the coordinates of the Wigner function $W(q,p)$: a GKP state with smaller $\Delta$ has more photons, incurring a larger error probability due to photon loss.
Based on the interplay between these two mechanisms, Appendix S$4.1$ in~\cite{CI+:grid} suggests that $\sigma=\frac{1}{2} \sqrt{\frac{\kappa_c T}{2}}$, with $T$ the total duration of the $S_p$ and $S_q$ measurement protocol, is a target value for squeezing (in our convention corresponding to $\Delta=\frac{1}{2}\sqrt{\kappa_c T}$).
A shorter cycle time $T$ can thus allow for a smaller $\Delta$, leading to a GKP qubit with a lower logical error rate.

We can compare our scheme with the proposed fault-tolerant syndrome detector of a GKP qubit in~\cite{puri+:detector-GKP}.
In that paper, it is proposed that a Kerr-cat qubit with $\ket{0} \approx \ket{\alpha}$ and $\ket{1} \approx \ket{-\alpha}$, is used for sequentially extracting bits of phase information of $S_q$ instead of a transmon ancilla qubit as in~\cite{CI+:grid}.
The advantage of using a Kerr-cat qubit is that unlike the transmon qubit or the scheme proposed here, there is little feedback error since the $X$-error rate on the Kerr-cat qubit is purposefully low, with photon loss leading only to $Z$-errors which do not feedback.
Note also that in~\cite{puri+:detector-GKP} the required coupling between the Kerr-cat qubit and the target (GKP) oscillator is not directly a photon-pressure coupling but a tunable beamsplitter interaction $\sim a^{\dag} b+a b^{\dag}$.

\subsection{Readout Loss}
\label{sec:read-out}
After the interaction of the target and the ancilla oscillator ---during the release and heterodyne measurement of the state of the ancilla oscillator--- one expects losses, and possibly thermalization,
due to coupling to extraneous modes in the co-planar or co-axial waveguide, circulators or the amplifier, affecting the total coherent amplitude of the ancilla oscillator state to be read out.
Since these losses result from various (partially unknown) sources, it is most reasonable to model these processes as a phenomenological loss process mapping the coherent amplitude $\alpha$ onto $\alpha_{\rm eff} < \alpha$, i.e. $U_{\rm loss}\ket{\alpha}_A\ket{0}_{\rm env} \rightarrow \ket{\cos(\theta) \alpha}_A \ket{\sin(\theta)\alpha}_{\rm env}$ with $\cos^2(\theta)\alpha^2=\alpha_{\rm eff}^2$ where $\ket{\gamma}_{\rm env}$ is some environment mode. We thus assume that these losses do not further influence the phase of the state $\alpha$.
The cumulative effect of losses is not expected to be small, for example in~\cite{pfaff+:controlled-release} $\eta_{\rm eff} ={(\alpha_{\rm eff}/\alpha)}^2 \approx 0.43$.

The effect of these losses is that some of the information about $S_q$ ends up in the environment and is not observed, leading to noise. We can simply modify the analysis in \cref{sec:toy} by inserting $U_{\rm loss}$ after $U_{\rm PP}$ of \cref{eq:U_{TA}} and prior to the heterodyne measurement action with outcome $\beta$, tracing over the environment mode. We get
\begin{equation}
\rho_{\rm in} \rightarrow \rho_{\beta}=\frac{1}{\pi} \int \diffd q\; \int \diffd q'\; \langle \alpha S_{q'} \sqrt{1-\eta_{\rm eff}} |\alpha S_q
 \sqrt{1-\eta_{\rm eff}}\rangle \times \langle \beta|\alpha_{\rm eff} S_q \rangle \langle \alpha_{\rm eff}S_{q'} | \beta \rangle \bra{q} \rho_{\rm in}\ket{q'}  \ket{q}\!\bra{q'},
\end{equation}
where $S_q$ and $S_{q'}$ are understood to be phases not operators.
Let us again analyze the two possible effects of loss.
First, for the diagonal elements of $\rho_{\rm in}$ in the $\ket{q}$-basis, the effect of the measurement is to apply the measurement operator $M_{\beta}(\alpha_{\rm eff})$.
Since the expected value for $\Delta_q$ only depends on the diagonal elements $\bra{q} \rho_{\rm in} \ket{q}$, this results in a higher expected value for $\Delta_q$ simply due to $\alpha \rightarrow \alpha_{\rm eff}$:
it is as if one executes the $S_q$-measurement with a smaller coherent state with amplitude $\alpha_{\rm eff}$.
Secondly, is there additional dephasing effect in the $q$-basis? Note that the measurement with subsequent loss in the ancilla oscillator still commutes with the operator $X$ or $S_p$, similar as the ideal measurement that we examined previously.
This directly means that $|{\rm Tr} S_p \int \diffd\beta\; \rho_{\beta}|=|{\rm Tr} S_p \rho_{\rm in}|$ and the same for $X$, \ie the average state has the same sharpness. 
We can also examine the sharpness averaged over different outcomes, that is, $\int \diffd\beta\; |{\rm Tr} X \rho_{\beta}|$.
Using that $\bra{q} X \ket{q'}=\bra{q} q'+\sqrt{\pi}\rangle=\delta(q-q'-\sqrt{\pi})$ and $\langle \alpha S_{q'} \sqrt{1-\eta_{\rm eff}} |\alpha S_q \sqrt{1-\eta_{\rm eff}}\rangle \delta(q-q'-\sqrt{\pi})=1$, the latter expressing the commutation of $S_q$ with $X$ (or $S_p$) we can write
\begin{equation}
\int \diffd\beta\; |{\rm Tr} X \rho_{\beta}|=\int \diffd\beta\; \left| \int \diffd q\; |\langle \beta|\alpha_{\rm eff} S_q \rangle|^2   \bra{q} \rho_{\rm in} \ket{q-\sqrt{\pi}}\right|=\int \diffd \beta  \int \diffd q\;  |\langle \beta|\alpha_{\rm eff} S_q \rangle|^2   \bra{q} \rho_{\rm in} \ket{q-\sqrt{\pi}}={\rm Tr} {\rm X} \rho_{\rm in},  
\end{equation}
whenever $\bra{q} \rho_{\rm in} \ket{q-\sqrt{\pi}} \geq 0$. Similarly, when $\bra{q} \rho_{\rm in} \ket{q-2\sqrt{\pi}} \geq 0$,  the mean sharpness $\int d\beta| {\rm Tr} S_p \rho_{\beta}|={\rm Tr} S_p \rho_{\rm in}$, is also unchanged by the $S_q$-measurement. These conditions are clearly fulfilled for the vacuum state.

The upshot of these considerations is that loss further down in the measurement chain only changes the effective strength of the coherent state that is used: when losses are such that $\eta_{\rm eff}=50\%$ and we use $\bar{n}=4$, we effectively get the squeezing due to $\bar{n}=2$, but no other extra noise.

\subsection{Third-order Nonlinearity}
\label{sec:nl}

\begin{figure}[htb]
	\includegraphics[width=\textwidth]{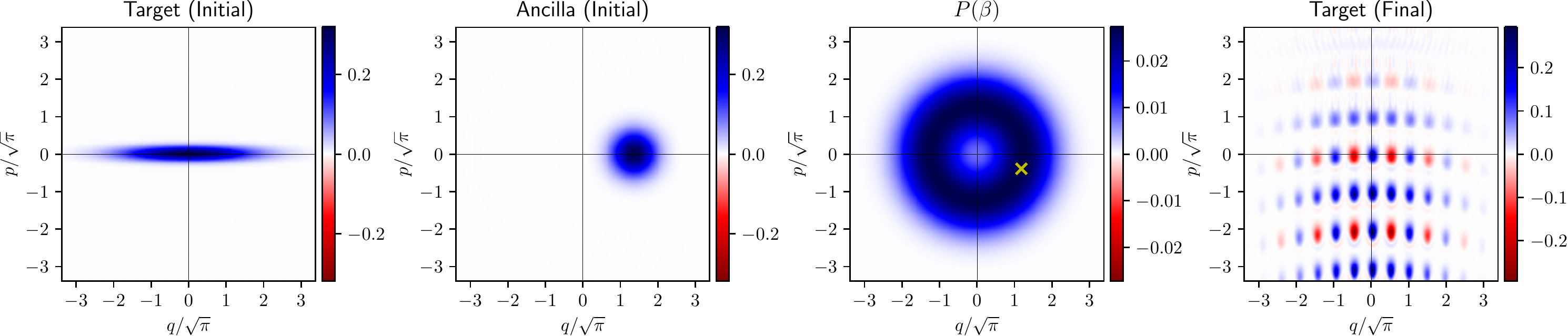}\\
	\ \\ 
	\includegraphics[width=\textwidth]{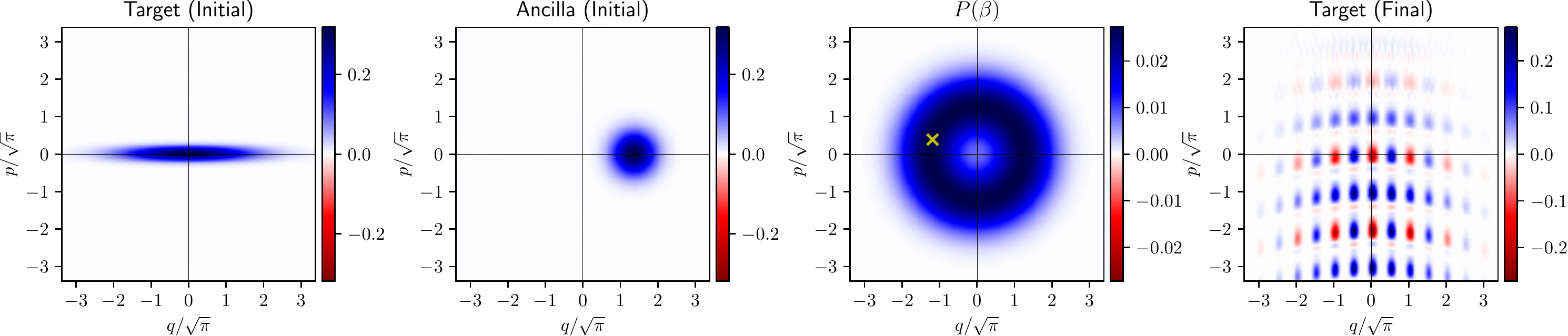}
	\caption{Wigner functions and probability distribution $\mathbb{P}(\beta)$ over measurement results using the heterodyne measurement of the ancilla oscillator, including the leading nonlinear term.
	The initial states are a squeezed vacuum state with $\Delta_q=3$ and $\Delta_p=1/3$ in the target oscillator and the coherent state $\ket{\alpha=\sqrt{3}}$ in the ancilla oscillator. The measurement result is the one with the maximum likelihood with respect to $\mathbb{P}(\beta)$.
	The strength of the third order term is set to $\frac{\xi_T^2}{\xi_A^2}=10^{-3}$, compare \cref{tab:properties}.
	Top: Original third-order nonlinearity according to \cref{eq:U3}. The effective squeezing of the final state is $\Delta_p=0.42, \Delta_q=0.2$.
	Bottom: Third-order nonlinearity with a modified drive, see \cref{eq:U3_mod}. The effective squeezing of the final state is $\Delta_p=0.41, \Delta_q=0.18$, demonstrating that $\Delta_q$ is unchanged compared to the ideal measurement in Fig.~\ref{fig:wigner-simple}.
	}
\label{fig:wigner-nonlinear}
\end{figure}

In this section we examine what happens when we include a leading-order correction in the Hamiltonian obtained from the circuit analysis from \cref{sec:circuit}.
The most important term neglected in the circuit analysis is $\propto \xi_T^3{\left(b^\dag\e^{\i\omega_T t}+b\e^{-\i\omega_T t}\right)}^3$ as this term is partially made resonant by the drive $x_{\rm ext}(t)$.

In this section we will see that it is crucial that any device fulfills $\xi_T^2/\xi_A^2 \ll 1$ because this ratio sets the strength of the unwanted unitary operation due to the leading-order correction.

We start with the original Hamiltonian in the rotating frame \cref{eq:H_RWA_orig}, but also keep the terms in the $\sin()$ part which oscillate with frequency $3\omega_T$ for now:
\begin{align}
	H_{\rm RWA,3} &\approx
	E_J \cos(x_{\rm ext}(t))
	\left(
		\frac{\xi_A^2 \xi_T^2}{2}(a^\dag a + b^\dag b + 2 a^\dag a b^\dag b)
		+\frac{\xi_A^4}{4}(a^\dag a + {(a^\dag a)}^2)
		+\frac{\xi_T^4}{4}(b^\dag b + {(b^\dag b)}^2)
	\right)\notag \\
	&+E_J \sin(x_{\rm ext}(t))
	\left(\xi_T
		\left(1-\frac{\xi_A^2}{2}-\xi_A^2 a^\dag a\right)(b^\dag \e^{\i\omega_T t} + b \e^{-\i\omega_T t})-\frac{\xi_T^3}{6}{\left(
			b^\dag\e^{\i\omega_T t}+b\e^{-\i\omega_T t}\right)}^3
	\right).
	\label{eq:H_modified}
\end{align}
Next, we use the properties of the drive $x_{\rm ext}(t)$ discussed in \cref{sec:drive}, \ie we drop all terms proportional to $\cos(x_{\rm ext}(t))$ and use $\sin(x_{\rm ext}(t))=\cos(\omega_T t)$:
\begin{align}
	H_{\rm RWA,3} &\approx
	\frac{E_J}{2}
	\left(\xi_T
		\left(1-\frac{\xi_A^2}{2}-\xi_A^2 a^\dag a\right)(b^\dag + b) -\cos(\omega_T t)\frac{\xi_T^3}{3}{\left(
			b^\dag\e^{\i\omega_T t}+b\e^{-\i\omega_T t}\right)}^3
	\right). \notag
\end{align}
If we expand the product ${\left(b^\dag\e^{\i\omega_T t}+b\e^{-\i\omega_T t}\right)}^3$, we see that all terms except $b^3\e^{-3\i\omega_T t}$ and ${(b^\dag)}^3\e^{3\i\omega_T t}$ oscillate with frequency $\omega_T$, compare \cref{eq:H_RWA}.
If we now use $\cos(\omega_T) = \frac{1}{2}(\e^{\i\omega_T t}+\e^{-\i\omega_T t})$ and drop all time dependent terms in the Hamiltonian,
we have
\begin{align}
	\tilde{H}_{on} &\approx
	\frac{E_J}{2}
	\left[
		-\xi_T \xi_A^2 a^\dag a(b^\dag +b)
		+\frac{\xi_T^3}{6}({(b^\dag+b)}^3-{(b^\dag)}^3-b^3)
	\right]\notag\\ 
	&=
	\frac{E_J}{\sqrt{2}}
	\left[
		-\xi_T \xi_A^2 a^\dag a \hat{q}
		+\frac{\xi_T^3}{6\sqrt{2}}(2\sqrt{2}\hat{q}^3-{(b^\dag)}^3-b^3)
	\right]  \notag.
\end{align}
This Hamiltonian acts for a fixed time $t_{\rm coupl} = \frac{2\sqrt{2\pi}}{E_J\xi_T\xi_A^2}$. If we also drop the unconditional displacement, the target and ancilla oscillators are coupled by the modified unitary operator
\begin{align}
	\tilde{U}_{\rm PP} = \exp\left(\i
		2\sqrt{\pi} a^\dag a \hat{q}
		+\i \frac{\sqrt{\pi}}{3\sqrt{2}}\frac{\xi_T^2}{\xi_A^2}\left(2\sqrt{2}\hat{q}^3-b^3-{(b^\dag)}^3\right)
		\right).
		\label{eq:U3}
\end{align}
Using $\epsilon=\frac{\sqrt{\pi}\xi_T^2}{3\xi_A^2\sqrt{2}}$, we can rewrite and approximate this unitary as
\begin{eqnarray}
\tilde{U}_{\rm PP} & \approx &  U_{\rm PP}\exp(\i 2 \sqrt{2} \epsilon \hat{q}^3)
\exp\left(-\i \epsilon(b^3+{(b^{\dag})}^3)\right) \exp\left(\sqrt{\pi} \epsilon a^{\dagger} a [ \hat{q}, (b^3+{(b^{\dag})}^3)]\right) \nonumber \\
& = & U_{\rm PP}\exp(\i 2 \sqrt{2} \epsilon \hat{q}^3)
\exp\left(-\i \epsilon(b^3+{(b^{\dag})}^3)\right) \exp\left(\frac{3\sqrt{\pi}\epsilon }{\sqrt{2}}  a^{\dagger}a \,({(b^\dag)}^2-b^2)\right).
\label{eq:nl-effect}
\end{eqnarray}
where we have neglected the commutators $\propto \epsilon^2$ and used that $[\hat{q},b^3 + {(b^\dag)}^3] = \frac{3}{\sqrt{2}} ({(b^\dag)}^2-b^2)$.
We observe two effects. First, the incorrect unitary induces a systematic (third-order) error of strength $\sim \epsilon$ on the target oscillator, independent of the ancilla oscillator, hence not affecting the outcome of the $S_q$ measurement itself. This systematic error does however cause a deformation of the Wigner function of a GKP code state.
Namely, if one applies to an approximate GKP state a unitary of the form $\exp(\i \delta q^3)$ with some parameter $\delta$ it will not change its squeezing $\Delta_q$, but it does lead to enhanced $\Delta_p$. Also, if we apply to a GKP state a unitary of the form $V=\exp( \i \delta (b^3+{(b^{\dag})}^3))$, it negatively affects the squeezing $\Delta_q$ as $V$ does not commute with $S_q$. Both effects are more pronounced the more photons the GKP state has.

Secondly, we observe that \cref{eq:nl-effect} contains an additional coupling between target and ancilla oscillator of the form $\exp(\delta a^{\dagger} a ({(b^{\dagger})}^2-b^2))$\footnote{If we had kept the unconditional displacement interaction in Eq.~(\ref{eq:H_eff}), we would also get some ancilla oscillator independent squeezing.}.
We can see this as squeezing induced by the ancilla oscillator on the target oscillator which gets stronger the more photons the ancilla oscillator contains. Alternatively, the heterodyne measurement statistics will be slightly altered by the presence of this additional term.

This photon-number dependent squeezing limits the number of photons that can be used in the ancilla oscillator. To alleviate this issue and ensure that the effective squeezing $\Delta_q$ is unchanged, one could apply a modified, two-tone drive such that
$\sin(x_{\rm ext}(t))\approx\cos(3\omega_T)+\cos(\omega_T)$, which has the effect of making $b^3$ and ${(b^{\dag})}^3$ terms resonant again.
Note that the terms $b^3$ and ${(b^{\dag})}^3$ are the only ones in the coupling Hamiltonian \cref{eq:H_modified} oscillating at frequency $3\omega_T$, the next highest order affected by the modified drive it is the fifth order of the expansion.
With the modified drive, the unitary time evolution only depends on $q$:
\begin{align}
\tilde{U}_{\rm PP}^{\rm corr} = \exp\left(\i
	2\sqrt{\pi} a^\dag a \hat{q}
	+\i 2\sqrt{2}\epsilon\hat{q}^3
	\right)=U_{\rm PP} \exp\left(\i 2\sqrt{2}\epsilon\hat{q}^3
	\right).
	\label{eq:U3_mod}
\end{align}
This corrected unitary transformation will then not affect the measurement statistics of $S_q$ as the additional term commutes with $\hat{q}$. The effective squeezing $\Delta_q$ of the measured state will be unchanged as compared to using $U_{\rm PP}$. The effective squeezing $\Delta_p$ is still affected by the $\exp(\i \delta \hat{q}^3)$ deformation.
The deformation can be seen as a displacement that has a quadratic dependence on the $\hat{q}$ quadrature, leading to a `parabola' of displacements acting on the final state of the target oscillator, see \cref{fig:wigner-nonlinear}.

The upshot is that with additional drive engineering one can mitigate the effect of the third-order non-linearity. The numerics in Fig.~\ref{fig:wigner-nonlinear}  show that for sufficiently small corrections the effect on the squeezing parameters is moderate.

\subsection{Flux Noise}
\label{sec:fn}
Because an external flux drive is used to enable the coupling, the setup will be susceptible to flux noise.
(Quasi)-static flux noise acts as a constant offset on the drive in \cref{eq:drive}.
Thus, with a constant flux offset $\epsilon$, \ie $\tilde{x}_{\rm ext}(t)=x_{{\rm ext},\pm}(t)+\epsilon$ with $x_{{\rm ext},\pm}(t)$ chosen as in Eq.~(\ref{eq:drive}) and maximal coupling strength ($\delta=1$), the interaction Hamiltonian (in the rotating frame) is given by
\begin{align}
	\tilde{H}_{\rm RWA}
	&\approx
	E_J \cos(\epsilon \pm {(-1)}^{k}\omega_T)
	\left(
		\xi_T (1-\frac{\xi_A^2}{2})(b^\dag \e^{\i\omega_T t} + b \e^{-\i\omega_T t})
		-\xi_T \xi_A^2 a^\dag a(b^\dag \e^{\i\omega_T t} + b \e^{-\i\omega_T t})
	\right) \notag\\ 
	&= \frac{E_J}{2}
	\left(
	\xi_T (1-\frac{\xi_A^2}{\sqrt{2}})(
		\hat{q} \cos(\epsilon) \mp {(-1)}^{k} \hat{p} \sin(\epsilon)
	)
		-\xi_T \xi_A^2 a^\dag a(
			\hat{q} \cos(\epsilon) \mp {(-1)}^{k} \hat{p} \sin(\epsilon)
		)
	\right),\notag
\end{align}
where the sign $\pm$ depends on the chosen drive and $k=\lfloor \frac{\omega_T t}{2\pi}\rfloor$ indicates the number of periods $\omega_t/(2\pi)$ that has passed by the time $0<t<t_{\rm coupl}$.
This Hamiltonian is still of the photon-pressure type, but it no longer couples the $\hat{q}$ quadrature to the number of photons in the ancilla oscillator, but a slightly rotated quadrature.
However, we can also see that this rotation is time-dependent due to its dependence on $k$ and changes direction with frequency $\omega_T$.
This means that the drive \cref{eq:drive} already provides some protection against such static flux noise.
In the case where a drive with reduced amplitude $\delta<1$ (see \cref{sec:drive}) is used, the situation is more complicated. We discuss flux noise for $\delta<1$ in \cref{sec:flux_delta}.

Another effect of flux noise is the following. The resonance frequency of both oscillators also depends on the external flux drive $x_{\rm ext}(t)$, see \cref{sec:circuit,tab:properties}.
In the presence of static flux noise, it means that the rotating frame will be slightly out of sync with respect to the true resonance frequency of the oscillators, leading to inaccuracy in the phase of the oscillator state. Typically, flux noise is small compared to $\Phi_0$ (which is the amplitude of the flux drive $x_{\rm ext}(t)$), suggesting that the difference between the expected and true resonance frequencies can be neglected.


\section{Discussion and Acknowledgements}
\label{sec:discuss}

In this paper we have proposed to use a simple coherent state ancilla to get more than 1 bit of information about the eigenvalue of a unitary displacement operator, effectively realizing a modular quadrature measurement. These measurements can be used to prepare or read out a GKP code state.
We have presented and analyzed an electric circuit which generates a strong photon-pressure coupling needed to imprint the eigenvalue information onto the coherent state of the ancilla oscillator.
The photon-pressure coupling Hamiltonian realized by this circuit is very versatile, with a simple modification of the flux drive, it can also be used to enable a beam-splitter between the target and ancilla oscillators~\cite{AKM:RMP, Eichler.Petta.2017:CqedOptomechanicsExp}.

As we have seen a large coherent amplitude $\alpha$ makes for possibly higher-precision stabilizer measurement, but in the presence of photon loss or unwanted nonlinearities, $\alpha$ should be chosen moderately. Our results and numerics show that circuit parameters can be chosen which demonstrate good performance at $|\alpha|^2 \approx 3$.

Our work was supported by ERC grant EQEC No. 682726. We thank Alexandre Blais, Daniel Bothner, Alessandro Ciani, Jonathan Conrad, Ines Corveira Rodrigues, David DiVincenzo and Wolfgang Pfaff for feedback and discussions.

\appendix
\section{Expansion of The Circuit Hamiltonian}
\label{sec:expansion}
During the circuit analysis in \cref{sec:circuit}, we expand the potential part of the circuit Hamiltonian \cref{eq:pot} around the approximate minimum $\mathbf{x}_A=\mathbf{x}_T=0$ of the potential term. In this section, we discuss this approximation in more detail as this point is not exactly the minimum of the potential.

Although the minimum of \cref{eq:pot} is not soluble analytically, we can find an upper bound on the errors made. We do this by investigating the maximal possible shift in the position of the minimum as a function of $x_{\rm ext}(t)$.
First, we expand the potential exactly using the addition formula of the cosine:
\begin{align*}
	U(\mathbf{x}_A,\mathbf{x}_T)
	&=\frac{E_{L_A}\mathbf{x}^2_A}{2}
		+\frac{E_{L_T}\mathbf{x}^2_T}{2}
		-E_J \cos\left(\mathbf{x}_T -\mathbf{x}_A-x_{\rm ext}(t)\right)\\
	&=\frac{E_{L_A}\mathbf{x}^2_A}{2}
		+\frac{E_{L_T}\mathbf{x}^2_T}{2}
		-E_J \cos(x_{\rm ext}(t))\cos\left(\mathbf{x}_T -\mathbf{x}_A\right)
		+E_J \sin(x_{\rm ext}(t))\sin\left(\mathbf{x}_T -\mathbf{x}_A\right).
\end{align*}
Because $E_J<E_{L_T},E_{L_A}$, we can see that the potential always has a unique minimum, and because the cosine is an even function, the location of that minimum only depends on the sine part of the equation.
Therefore, the maximal shift of the position of the minimum away from $\mathbf{x}_T=\mathbf{x}_A=0$ occurs at $x_{\rm ext}(t)=\pm \pi/2$.

If we set $x_{\rm ext}=\pi/2$ and expand the potential to first order around $\mathbf{x}_A=\mathbf{x}_T=0$, we have
\begin{align}
	U(\mathbf{x}_A,\mathbf{x}_T)
	&\approx\frac{E_{L_A}\mathbf{x}^2_A}{2}
		+\frac{E_{L_T}\mathbf{x}^2_T}{2}
		+E_J \left(\mathbf{x}_T -\mathbf{x}_A\right)\notag\\ 
	&=\frac{E_{L_A}}{2}
	{\left(\mathbf{x}_A-\frac{E_J}{E_{L_A}}\right)}^2
	+\frac{E_{L_T}}{2}
	{\left(\mathbf{x}_T+\frac{E_J}{E_{L_T}}\right)}^2
	-\frac{E_J^2}{2E_{L_T}}-\frac{E_J^2}{2E_{L_A}}.
	\label{eq:U_minimum}
\end{align}
Thus, for $x_{\rm ext}(t)=\pm \pi/2$, the minimum of the potential term \cref{eq:pot} is located around $\mathbf{x}_T = \mp \frac{E_J}{E_{L_T}}$ and $\mathbf{x}_T = \pm \frac{E_J}{E_{L_T}}$.
By defining $\mathbf{x}_T' = \mathbf{x}_T\pm \frac{E_J}{E_{L_T}}$, $\mathbf{x}_A' = \mathbf{x}_A\pm \frac{E_J}{E_{L_T}}$ we can absorb this correction into the external flux drive $x_{\rm ext}(t)$
\begin{align*}
	U(\mathbf{x}_A',\mathbf{x}_T',x_{\rm ext}(t)=\pm\pi/2)
	&=\frac{E_{L_A}\mathbf{x'}^2_A}{2}
		+\frac{E_{L_T}\mathbf{x'}^2_T}{2}
		-E_J \cos\left(\mathbf{x}_T' -\mathbf{x}_A'\pm2\frac{E_J}{E_{L_T}}-\mp \pi/2\right),
\end{align*}
where the minimum of the potential is now to first order given by $\mathbf{x}_A'=\mathbf{x}_T'=0$.
Because the sine is monotone between $0$ and $\pi/2$, we know that the location of the true minimum of the potential is also monotone between $0<x_{\rm ext}(t)<\pi/2$.
From the structure of \cref{eq:U_minimum} we can also see that the shift of the minimum for $\mathbf{x}_T$ is also always opposite to that of $\mathbf{x}_A$ and that the sign of this minimum changes for $-\pi/2<x_{\rm ext}(t)<0$.
\begin{align*}
U(\mathbf{x}_A',\mathbf{x}_T')
	&=\frac{E_{L_A}\mathbf{x'}^2_A}{2}
		+\frac{E_{L_T}\mathbf{x'}^2_T}{2}
		-E_J \cos\left(\mathbf{x}_T' -\mathbf{x}_A' +2\epsilon+x_{\rm ext}(t)\right),
\end{align*}
where $|\epsilon| \lesssim \frac{E_J}{E_{L_T}}$ is the true location of the minimum of $U(\mathbf{x}_A',\mathbf{x}_T')$ and the sign of $\epsilon$ depends on the sign of $\sin(x_{\rm ext}(t))$.
The expansion around $\mathbf{x}_A=\mathbf{x}_T=0$ made in \cref{sec:circuit} is therefore similar to the effect of oscillating flux noise.
Furthermore, just like for flux noise, there is an echo effect reducing any contributions from this offset because the offset changes sign with frequency $\sim \omega_T$.

In summary, the problem that arises is that the external drive changes the potential for both variables, which follow the change of minimum, with some delay.
Due to this delay, the instantaneous potential is not quite what we expect, but the error is small, as it scales as $E_J/EL \ll 1$.
Note that this effect is deterministic, so it could be counteracted by a change in the external drive.

\section{Details of the Flux Drive}
A key component to achieve a photon-pressure coupling in the rotating frame is a suitable flux drive that cancels the time-dependence of a Hamiltonian in the rotating frame.
In the following subsections we discuss how this drive can be achieved, discuss details on flux noise and show how a microwave drive could be used instead of a flux drive.

\subsection{Parametric Flux Drive}
\label{sec:drive}
\begin{figure}
	\includegraphics[width=0.4\textwidth]{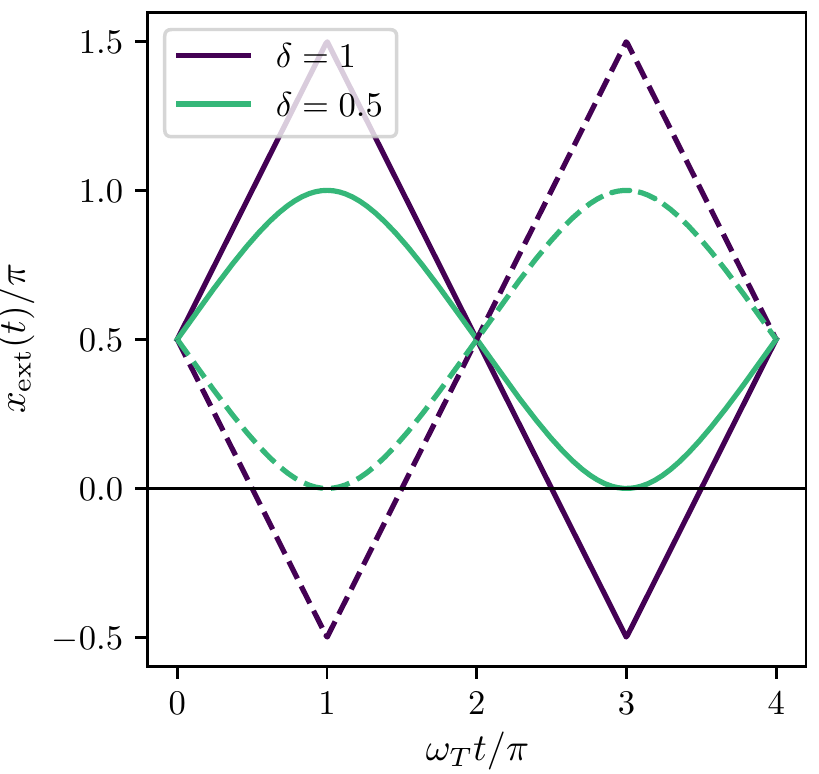}
	\hfill
	\includegraphics[width=0.4\textwidth]{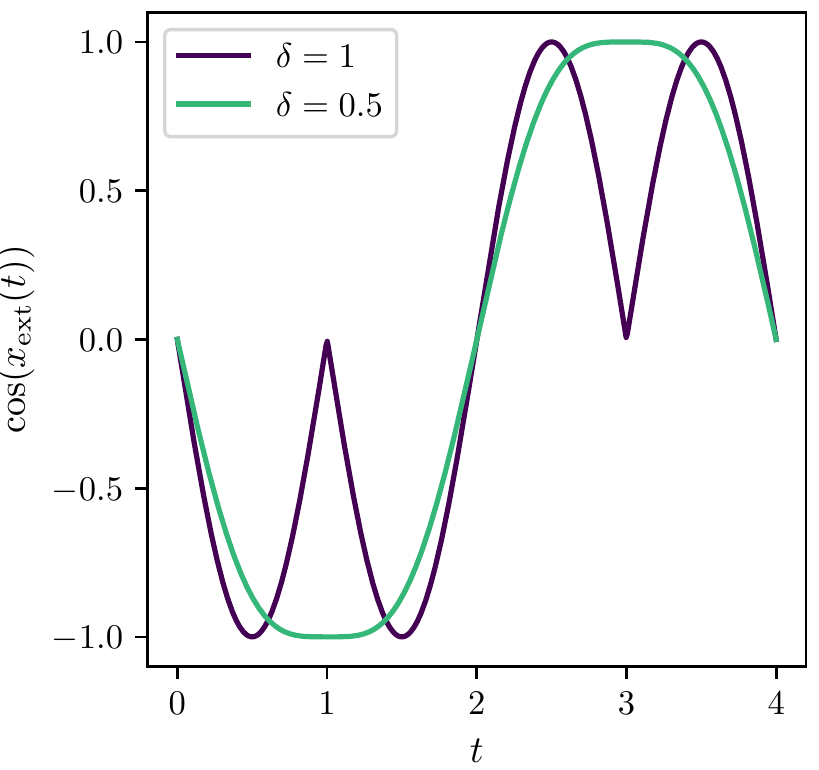}
	\caption{
		Left: The external drive $x_{\rm ext}(t)$ required to obtain a photon-pressure coupling in the rotating frame, see \cref{eq:drive}. As can be seen, the lowest frequency component of $x_{\rm ext}(t)$ is $\omega_T/2$. The starting point of $x_{\rm ext}(t=0)=\pi/2$ corresponds to the ``off'' setting where the target and ancilla oscillators are completely decoupled.
		The solid and dashed lines correspond to choosing the drive with a positive or negative sign, $x_{{\rm ext},\pm}(t)$, respectively.
		Right: The prefactor $\cos(x_{\rm ext}(t))$ of the self- and cross Kerr terms in \cref{eq:H_RWA_orig}. The function is periodic and changes sign with frequency $\omega_T$.\\
		Purple: Drive required to obtain the maximum coupling strength \ie $\delta=1$. The drive corresponds to a triangular wave.
		Green: the coupling strength is reduced to $\delta=0.5$. In this case, the drive is close to a simple cosine.
	}
	\label{fig:drive}
\end{figure}


To achieve the desired photon-pressure coupling from \cref{eq:H_desired}, it is necessary to design an appropriate time-dependence of $x_{\rm ext}(t)$ in \cref{eq:H_RWA} such that the phases $\e^{\pm\i\omega_T t}$ in that equation cancel.
The idea is similar to the case of qubit readout,~\cite{touzard:displace, Ikonen:meas,DBB:fast-para}, but here we can use that the frequency of the target oscillator is relatively small in order to maximize the coupling strength, which is not the case for qubit readout.
Furthermore, we can use a flux drive with an amplitude of $2\pi$, cancelling the anharmonicity of both oscillators (something which is undesired in the case of qubit readout).

\begin{figure}
	\includegraphics[width=0.4\textwidth]{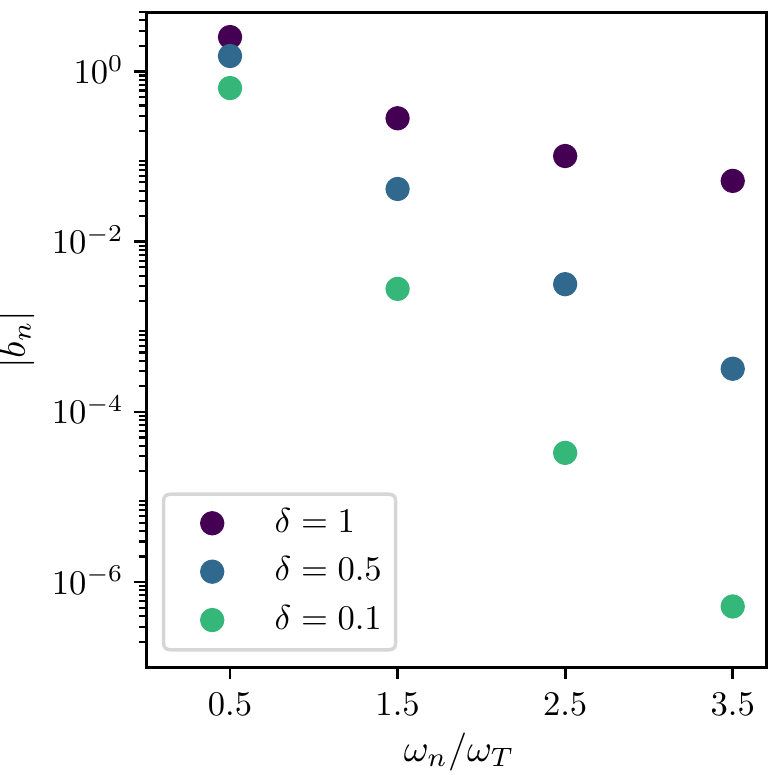}
	\caption{
		Amplitude $|b_n|$ of the first four harmonics with frequency $\omega_n$ of the Fourier series of either drive $x_{\rm ext, \pm}(t)$, see \cref{eq:drive}. Purple: Maximal coupling strength ($\delta=1$), Blue: $\delta=0.5$, Green $\delta=0.1$.
		For $\delta=0.5$, it is sufficient to use only two harmonics in order to achieve a relative error below $1\%$.}
	\label{fig:harmonics}
\end{figure}

To this end, we consider a flux drive such that $\sin(x_{\rm ext}(t)) = (1-\delta)+\delta\cos(\omega_T t)$, where $0<\delta<1$ is a freely chosen constant which serves to reduce the amplitude of the flux drive.
Scenarios where this is desirable are, for example, if a lossy resonator is used to implement the tunable coupling, or if the range of resonance frequencies should be limited.
One can easily verify that either drive
\begin{align}
	x_{\rm ext, \pm}(t)= \frac{\pi}{2}
		\pm{(-1)}^{\left\lfloor \frac{\omega_T t}{2\pi} \right\rfloor}\arcsin(1-\delta+\delta\cos(\omega_T t))
		\label{eq:drive}
\end{align}
satisfies that condition.
For $\delta=1$, we can also see that $\cos(x_{\rm ext, \pm}(t)) = \pm{(-1)}^{\left\lfloor \frac{\omega_T t}{2\pi} \right\rfloor}|\sin(\omega_T t)|$, corroborating the claim that the even order terms in \cref{eq:H_RWA} cancel.
Although this drive seems to be very complex, this function can be easily synthesized with a small number of harmonics.
In fact, the most complex possible drive (using the full flux range for maximal coupling strength, $\delta=1$) yields a triangular wave which rolls off with the inverse harmonic number squared:
\begin{align*}
	x_{\rm ext, +}(t) = \frac{\pi}{2}-\frac{8}{\pi}\sum_{n =0}^\infty \frac{{(-1)}^{n}}{{(2n+1)}^2}
	\sin\left(\frac{2n+1}{2}\omega_T t\right).
\end{align*}
Although the Fourier series of the drive does not have such a simple solution for $\delta<1$, it can be well approximated numerically, using $x_{\rm ext,\pm}(t)=\pi/2\pm\sum_{n=0}^\infty b_n \sin\left((2n+1)\omega_T t/2 \right)$.
The amplitude $|b_n|$ of the Fourier series of the drive is shown in \cref{fig:harmonics}.
As can be seen there, the roll off is fast, such that two harmonics are in many cases a sufficient approximation.
In addition, the period $\frac{4\pi}{\omega_T}$ of this drive is rather long, as the resonance frequency $\omega_T$ is typically in the regime $\sim 500$ MHz, see \cref{tab:properties}.
Due to the requirement that the resonance frequency of the ancilla oscillator should not exceed $\sim 10$ GHz, while there needs to be a separation of scales $\omega_A \gg \omega_T$ and $\omega_T$ should not be too small to avoid thermal excitations, this frequency range is not expected to change much for different setups.
As an estimate for the most complex case with $\delta=1$, the total error for a standard arbitrary waveform generator with $2.4$ Gsamples/sec without any corrections to the signal is expected to be around $0.5\%$.
Using either drive from \cref{eq:drive},
neglecting all terms rotating with frequency $\omega_T$ or above, the effective Hamiltonian from \cref{eq:H_RWA} yields the desired interaction:
\begin{align}
	H_{\rm on} \approx
	\frac{\delta}{2} E_J
		\left[
			 \xi_T (b^\dag + b)
			-\frac{\xi_T \xi_A^2}{2} (b^\dag +b)(2a^\dag a + 1)
		\right],
		\label{eq:H_eff-app}
\end{align}
with the coupling strength $g = \frac{\delta}{2}E_J \xi_T \xi_A^2$.

\subsection{Flux Noise with Reduced Drive Amplitude}
\label{sec:flux_delta}
We have discussed flux noise for the case with maximal coupling strength ($\delta=1$) in \cref{sec:fn}  of the main text.
Following the discussion there, we now discuss flux noise in the case where the drive strength is reduced ($\delta<1$). In this case the prefactor in the coupling Hamiltonian is given by
\begin{align}
\sin(x_{\rm ext, \pm}(t)+\epsilon) = (1-\delta + \delta \cos(\omega_T t)) \cos(\epsilon)
	\pm {(-1)}^{\lfloor \frac{\omega_T t}{2\pi}\rfloor} \sqrt{1-{(1-\delta+\delta\cos(\omega_T))}^2}\sin(\epsilon).
	\label{eq:drive_flux}
\end{align}
Again, there is some built-in correction for the additional phase, the term $\propto \sin(\epsilon)$ will approximately cancel over multiple periods (recall that the ideal drive is with zero flux off-set $\epsilon=0$).
However, there is an additional effect that the amplitude of the drive also changes over time, and the change of amplitude is on resonance with the change of the phase, \ie the rotation of the measured quadrature no longer completely cancels.
In order to alleviate this issue, one could use a similar strategy as the CZ gate used for transmon qubits~\cite{Rol.etal.2019:netzero}, using the fact that the equation $\sin(x_{\rm ext, \pm}(t)) = 1-\delta +\delta\cos(\omega_T t)$ has two alternating solutions, $x_{\rm ext,+}$ and $x_{\rm ext,-}$ see \cref{eq:drive}.
Because the undesired term due to flux noise for the two drives always has opposite sign, see \cref{eq:drive_flux},
it is possible to restore the echo effect by alternating between the two drives.
Although this transition is not smooth (see \cref{fig:drive}), it is continuous for all choices of $\delta$.
Furthermore, the quick roll off with high harmonics is preserved, the most complex drive that could be obtained with the strategy is a triangle wave at frequency $\omega_T$, see \cref{fig:drive}.


\subsection{Use of a Microwave Drive}
\label{sec:microwave_drive}
In most experimental settings, it is preferable to use a microwave drive instead of a time-dependent flux. Here, we show how such a microwave drive can be used, employing a similar method as Touzard \etal in~\cite{touzard:displace}.

Consider again the Hamiltonian from \cref{eq:H}, with the potential from \cref{eq:pot} and a classical field $y_{\rm ext}(t)\e^{\i\omega_d t}$ capacitively coupled to the ancilla oscillator.
Here, the phase $\e^{\i\omega_d t}$ indicates the lowest frequency term of the external drive, see \cref{sec:drive} for details. The system Hamiltonian is then given by
\begin{align}
	H_{CC} &= 4E_{C_A}\left(1-\frac{E_{C_A}}{E_{C_J}}\right)\mathbf{p}_A^2  + 4E_{C_T}\left(1-\frac{E_{C_T}}{E_{C_J}}\right)\mathbf{p}_T^2
	+\frac{8 E_{C_A}E_{C_T}}{E_{C_J}} \mathbf{p}_A \mathbf{p}_T
	+\frac{\i}{2}\mathbf{p}_A \Im(y_{\rm ext}(t)\e^{\i\omega_d t})\notag \\
&+\frac{E_{L_A}\mathbf{x}^2_A}{2}
	+\frac{E_{L_T}\mathbf{x}^2_T}{2}
	-E_J \cos\left(\mathbf{x}_T -\mathbf{x}_A-\pi/2 \right)\notag,
\end{align}
where the flux $x_{\rm ext}$ has been set to a constant value of $\pi/2$.
If we express this in terms of annihilation and creation operators, and collect all uncoupled quadratic terms into the harmonic part of the Hamiltonian, we have
\begin{align}
H_{CC} &= \omega_A a^\dag a + \omega_T b^\dag b
	-\frac{2E_{C_A}E_{C_T}}{E_{C_J}}\frac{1}{\xi_A\xi_T} (a^\dag - a)(b^\dag -b)
	-E_J \tilde{\cos}\left(\xi_T(b^\dag + b) -\xi_A(a^\dag+a)-\pi/2) \right) \notag\\ 
	&- \frac{1}{4\xi_A} (a^\dag  - a) (y_{\rm ext}^*(t)\e^{-\i\omega_d t}-y_{\rm ext}(t)\e^{\i\omega_d t}).
	\notag
\end{align}
Here, we use a notation analogous to Touzard \etal\cite{touzard:displace} and $\tilde{\cos}$ indicates that the second order terms $\xi_A^2{(b^\dag+b)}^2/2$, $\xi_T^2{(b^\dag+b)}^2/2$ of the cosine have already been absorbed in the harmonic part of the Hamiltonian.
Using the substitution $\tilde{a} = a + \frac{1}{4\omega_A \xi_A}y_{\rm ext}(t)\e^{\i\omega_d t}$, we get
\begin{align}
	H_{CC} &= \omega_A \tilde{a}^\dag \tilde{a} + \omega_T b^\dag b
	-\frac{2E_{C_A}E_{C_T}}{E_{C_J}} \frac{1}{\xi_A\xi_T}
	 \left(\tilde{a}^\dag - \tilde{a}+\frac{\i}{2\omega_A \xi_A} \Im\left(y_{\rm ext}(t)\e^{\i\omega_d t}\right)\right)(b^\dag -b)
	\notag\\ 
	&-\frac{1}{4\xi_A}\left(\tilde{a}^\dag y_{\rm ext}^*(t)\e^{-\i\omega_d t}+\tilde{a}y_{\rm ext}(t)\e^{\i\omega_d t}\right)
	-E_J \tilde{\cos}\left(\xi_T(b^\dag + b) -\xi_A(\tilde{a}^\dag+\tilde{a})-\pi/2-\frac{1}{2\omega_A}\Re(y_{\rm ext}(t)\e^{\i\omega_d t}) \right) \notag
\end{align}
The potential is again of the same form as \cref{eq:pot} (if we were to write $H_{CC}$ in terms of $\mathbf{x}_m,\mathbf{p}_m$ again), where the microwave drive $\Re(y_{\rm ext}(t)\e^{\i\omega_d t})/(2\omega_A)$ takes the role of the flux drive $x_{\rm ext}(t)$.
Note that there are two additional displacements, one acting on the target oscillator ($b^{\dag},b$) and one on the ancilla oscillator ($a^{\dag},a$). The drive on the target oscillator is an off-resonant, known, unconditional drive and can be corrected by a counterdrive.
If we go to a rotating frame and use the rotating wave approximation, the drive on the ancilla oscillator will vanish because it is very far off resonant (see \cref{sec:drive}).
Note that this step means that a microwave drive can only be used to obtain an oscillating drive, in order to obtain a constant offset, it is still necessary to use a (constant) flux drive, hence why we set $x_{\rm ext}=\pi/2$ in the beginning.
The term $\propto \left(\tilde{a}^\dag - \tilde{a}\right)\left(b^\dag -b\right)$ corresponds to the term $\mathbf{p}_A \mathbf{p}_T$ in \cref{eq:H} and will cancel in the rotating frame because the resonance frequencies of the two oscillators are very different.

\section{Heterodyne measurement via Release of Coherent Oscillator State into a Transmission Line}
\label{app:time-int}



Here we model the gradual release of the cavity state into a mobile wave packet traveling over a 1D transmission line or waveguide by an effective model.
Our goal is to verify that the integration of a heterodyne measurement signal on small coherent states released \emph{over time} can effectively give the same measurement operator as the direct heterodyne measurement of \cref{sec:toy}.
This is not immediately obvious. Given a long enough measurement time $t_{\rm meas}$, even if all photons in the oscillator are eventually measured to determine the angle, there are two combining features which could make such a measurement fundamentally more noisy than a direct heterodyne measurement and hence leading to less effective squeezing.
Firstly, the instantaneous measurement is applied to a \emph{small} coherent state, \ie the one that arrives during a small interval in time, which has \emph{large} angle uncertainty.
Secondly, the overall output of the measurement is only a weighted \emph{integration} of the heterodyne signal obtained from each small coherent state, \ie we assume that we gain no knowledge of the individual trajectory of outcomes, but only integrate (using a filter) their values in time, see \cref{eq:IQ} below (although one could go beyond this and look at full trajectories, see~\cite{hatridge+:meas,minev:jump}).
Note that in this effective model we do not include additional losses nor the amplification step in the measurement chain as we discuss their effect in \cref{sec:read-out}.
Naturally, due to the sequence of amplifiers and bringing the signal up to room temperature electronics, the actual states which are measured are not small coherent states, but classical voltage signals, but their quantum fluctuations are frozen in as classical noise.

In our analysis we also do not include a spurious photon-pressure coupling (and hence a possible rotation) to the target oscillator during the release of the state in the ancilla oscillator. Naturally, if the oscillator state is further rotated while it is also being released, then this is likely to lead to additional noise in the measurement as the to-be-measured-phase is changing in each of the weak measurements in sequence.

Our model is that of a sequence of $N$ beamsplitter interactions of strength $\sqrt{\delta t \times \kappa_{\rm open}}$ of the ancilla oscillator with individual `measure' modes $j=0,1,\ldots, J-1$ which are each initialized in a vacuum state.
For a non-tunable fixed set-up, the decay rate $\kappa_{\rm open}$ is determined by the capacitive coupling between ancilla oscillator and transmission line and enters a more complete Hamiltonian description of such an interaction, see \eg Appendix A in~\cite{BDT:meas}.
When one uses a $Q$-switch as in~\cite{pfaff+:controlled-release}, one can use an effective decay rate $\kappa_{\rm open}=4 g^2/\kappa_{\rm out}$ when $g \ll \kappa_{\rm out}$ where $g$ is the strength of the beam-splitter coupling to the lossy oscillator and $\kappa_{\rm out}$ is the decay rate of the lossy oscillator (set by its coupling strength to some transmission line/co-planar wave guide/co-axial cable hosting 1D continuous traveling modes).

The idea is that one has a beam-splitter interaction between oscillator mode and transmission line mode localized at a point in space:
this interaction puts some of the coherent amplitude in this spatial mode which due to the transmission-line Hamiltonian propagates away at (speed of light) velocity $v$, returning the local spatial mode to the vacuum, see Appendix E, Section 2 pp. 73--77 in~\cite{RMP:clerk+} for this perspective of the interaction of a (cavity) oscillator with the bath modes on the transmission line.
Hence the measure mode $j$ will model the state that one can measure at time $t=j \delta t$ at a fixed spatial point on the transmission line where the detector sits: a new measure mode is arriving at the detector after each time-step $\delta t$. We will take the continuum limit $\delta t \rightarrow 0$ and $J \rightarrow \infty$ in our expressions while keeping the total measurement time $t_{\rm meas}=\delta t J$ finite.
Note that we could include thermal noise in this model by having each measure mode initialized in a thermal state instead of a vacuum state.
We will assume that each measure mode $j$ undergoes a complete heterodyne measurement, providing an outcome $\beta_j$. In addition, we omit any time-dependence of the ancilla oscillator or the measure modes, \ie our expressions assume that we work in a rotating frame at the ancilla oscillator frequency.

The outcome of the measurement is an estimate of the time-integrated (dimensionless) quadratures $I_{\rm out}$ and $Q_{\rm out}$ which we define as
\begin{align}
&I_{\rm out}=\sqrt{2\kappa_{\rm open}} \int_0^{t_{\rm meas}} \diffd t\;  f(t)  \Re(\beta(t)), &&Q_{\rm out}=\sqrt{2\kappa_{\rm open}} \int_0^{t_{\rm meas}} \diffd t\; f(t)  \Im(\beta(t)).
\label{eq:IQ}
\end{align}
where $f(t)=\exp(-\kappa_{\rm open} t/2)$. Here $\beta(t)$ is the continuum limit of the outputs $\beta_j$, detailed below.
To make contact with the usual input-output formalism in which we have an outgoing field $b_{\rm out}(t)=\sqrt{\kappa_{\rm open}} a(t)$ for the cavity field $a$~\cite{RMP:clerk+} (represented here by the ancilla oscillator), we observe that the expected value
 $\langle I_{\rm out}\rangle=\int_0^{t_{\rm meas}} \diffd t\; \langle q_{\rm out}(t)\rangle $ where $q_{\rm out}(t)=\frac{1}{\sqrt{2}}(b_{\rm out}(t)+b_{\rm out}(t))$ (and similarly $\langle Q_{\rm out} \rangle=\int_0^{t_{\rm meas}} \diffd t\;  \langle p_{\rm out}(t) \rangle$).
The superoperator represented by this measurement is thus given as
\begin{eqnarray}
	{\cal S}_{Q_{\rm out},I_{\rm out}}(\rho_{\rm in})=\frac{1}{{\rm Tr}(\int_{Q_{\rm out},I_{\rm out}} \diffd\bm{\beta}\; M_{\bm{\beta}}^{\dagger} M_{\bm{\beta}} \rho_{\rm in})}\int_{Q_{\rm out},I_{\rm out}} \diffd \bm{\beta}\; M_{\bm{\beta}} \rho_{\rm in} M_{\bm{\beta}}^{\dagger},
\end{eqnarray}
where the integral goes over all $\bm{\beta}$ leading to integrated signal $Q_{\rm out}$ and $I_{\rm out}$. Based on $I_{\rm out}$ and $Q_{\rm out}$, the measurement estimates the eigenvalue $\exp(\i \varphi)$ of $S_q$ as $\varphi_{\rm out}=\arctan(Q_{\rm out}/I_{\rm out})$.
If we were to use a $Q$-switch and a lossy oscillator, the temporal profile of the outgoing field would not be the exponentially-decaying function $f(t)$ as the ancilla oscillator first has to build up some amplitude in the lossy oscillator before leaking out of it, and one could use such a compensated time-filter as in Eq. (S6) of~\cite{pfaff+:controlled-release}.

Now let us consider the details of this measurement. Our expressions will depend on $\kappa_{\rm open} t_{\rm meas}$ which we assume to be large, capturing the fact that we measure until the coherent state has entirely leaked out of the ancilla oscillator.
Each beamsplitter interaction $B$ applies a simple transformation on a coherent state $\ket{\beta}$ in the ancilla oscillator and a measure mode $j$:
\begin{equation}
B \ket{\beta}_A \otimes \ket{0}_j=\ket{\beta \cos(\sqrt{\kappa_{\rm open} \delta t})}_A \ket{\beta\sin(\sqrt{\kappa_{\rm open} \delta t})}_j\approx \ket{\beta \sqrt{1-\kappa_{\rm open}\delta t}}_A\ket{\beta\sqrt{\kappa_{\rm open} \delta t}}_j
\end{equation}
Let us write down the heterodyne measurement operator $M_{\bm{\beta}}$ for a sequence of outcomes $\beta_j$, $j=0,\ldots,J-1$, collectively denoted as a vector $\bm{{\beta}}$.
Note that the state of the ancilla oscillator and the measure modes after $J$ beamsplitters equals
\begin{align*}
	&B_{J-1}\dots B_2B_0\ket{\alpha}_A\ket{0}_{J-1}\dots\ket{0}_{2}\ket{0}_{0}
	= \ket{\alpha{(1-\kappa_{\rm open} \delta t)}^{J/2}}_A \prod_{j=0}^{J-1}\ket{\alpha_j}_{j},
&&\alpha_j \equiv \alpha{(\kappa_{\rm open} \delta t)}^{1/2}{(1-\kappa_{\rm open} \delta t)}^{j/2} \in \mathbb{R}.
\end{align*}
As the measure modes \(j=0,\dots, J-1\) do not couple, the total measurement operator on the measure modes is simply a product over all modes.
The measurement operator equals (using $\bra{\beta} \alpha \rangle=\exp(-(|\alpha|^2+|\beta|^2)/2) \exp(\beta^* \alpha)$):
\begin{align}
	M_{\bm{\beta}}&= \frac{1}{\pi^{J/2}}\prod_{j=0}^{J-1}\braket{\beta_j|\exp(\i 2\sqrt{\pi}\hat{q})\alpha_j}\nonumber \\
& = \frac{1}{\pi^{J/2}}\exp(-\frac{1}{2}\sum_{j=0}^{J-1} (|\alpha_j|^2+|\beta_j|^2))\exp\left(\sum_{j=0}^{J-1} \alpha_j \beta_j^* \exp(\i 2\sqrt{\pi}\hat{q})\right) \nonumber \\
&= \frac{1}{\pi^{J/2}}\exp(-\frac{1}{2}\sum_{j=0}^{J-1} (|\alpha_j|^2+|\beta_j|^2))\exp\left(\sum_{j=0}^{J-1} K_{|\beta_j|}\cos(2\sqrt{\pi}\hat{q}-\varphi_j)/2\right)\exp\left(\i \sum_{j=0}^{J-1} K_{|\beta_j|}\sin(2\sqrt{\pi}\hat{q}-\varphi_j)/2\right),
\label{eq:M_indirect}
\end{align}
using $K_{|\beta_j|}=2 \alpha_j |\beta_j|$. Not surprisingly, we see that the measurement operator has the same form as in \cref{eq:M_direct}.
If we take the continuum limit, we note that the $\hat{q}$-dependent part in $M_{\bm{\beta}}$ does not explicitly depend on the measurement results $\bm{\beta}$, but on a time-integrated average over the results as follows. We have
\begin{equation}
\sum_{j=0}^{J-1}\alpha_j \beta_j^*=  \sum_{j=0}^{J-1} \delta t \;\alpha \sqrt{\kappa_{\rm open}} {(1-\kappa_{\rm open}\delta t)}^{j/2} \frac{\beta_j^*}{
\sqrt{\delta t}}
\to  \int_0^{t_{\rm meas}}\diffd t\  \alpha(t) \beta^*(t),
\end{equation}
where we have defined $\beta(t)=\sqrt{\frac{\beta_j}{\delta t}}$ and $\alpha(t)=\alpha\sqrt{\kappa_{\rm open}}\e^{-\kappa_{\rm open} t/2}$. Note that $\alpha(t)$ and $\beta(t)$ have dimension $t^{-1/2}$.
Thus the $\hat{q}$-dependent part of $M_{\bm{\beta}}$ is, --in the continuum limit--, proportional to
\begin{equation}
\exp\left(\int_0^{t_{\rm meas}}\diffd t\; \alpha(t) \beta^*(t) S_q\right)=\exp\left(\frac{\alpha}{\sqrt{2}}(I_{\rm out}-\i Q_{\rm out}) S_q\right).
\end{equation}
Since $\sum_j |\beta_j|^2 \rightarrow \int_0^{t_{\rm meas}} dt |\beta(t)|^2$ and
$\sum_j |\alpha_j|^2 \rightarrow \kappa_{\rm open} |\alpha|^2 \int_0^{t_{\rm meas}} \diffd t\; \exp(-\kappa_{\rm open} t)=|\alpha|^2(1-\exp(-\kappa_{\rm open} t_{\rm meas}))\approx |\alpha|^2$, the prefactor in $M_{\bm{\beta}}$
does depend on $\int_0^{t_{\rm meas}} \diffd t\; |\beta(t)|^2$, not only on $Q_{\rm out}$ and $I_{\rm out}$.
The conclusion is that by using an exponentially-decaying filter on the measured data as in Eq.~\ref{eq:IQ}, one can ensure that a single measurement operator is applied on the input state given the measurement output $I_{\rm out}, Q_{\rm out}$ and this measurement operator does not depend on
the specific temporal noisy sequence $\beta_0,\ldots, \beta_{J-1}$. Hence we expect that the effect of this integrated measurement in time does not lead to a more noisy outcome than one in which we record the entire sequence of values $\beta_0,\ldots, \beta_{J-1}$.

We can make this explicit by estimating the effective squeezing as we have done in \cref{sec:toy} for the direct measurement. We can find
\begin{equation}
M_{\bm{\beta}}^{\dagger} M_{\bm{\beta}} \propto \exp\left(\alpha  \sqrt{2(Q_{\rm out}^2+ I_{\rm out}^2)}\cos(2\sqrt{\pi}\hat{q}-\varphi_{\rm out})\right),
\end{equation}
which defines an effective concentration $K_{\rm eff}=\alpha\sqrt{2(Q_{\rm out}^2+ I_{\rm out}^2)}$.
Hence in analogy with the direct measurement where the effective squeezing is estimated by considering $\langle |\beta| \rangle$, here the goal is to estimate the expected value of $K_{\rm eff}$. Translating back to the discrete representation, this requires estimating $\langle| \sum_j \alpha_j \beta_j^*| \rangle$. Instead of estimating $\langle| \sum_j \alpha_j \beta_j^*| \rangle$ we evaluate $\sqrt{\langle| \sum_j \alpha_j \beta_j^*|^2 \rangle} \propto \sqrt{\langle Q_{\rm out}^2+I_{\rm out}^2\rangle}$ and obtain a lower bound on the effective squeezing in this manner.

Using the discrete sequence-of-measurements representation, it can be observed that the entire measurement is a simple product of individual measurements each with outcome $\beta_j$ applied to a product state. We first observe that, like in the proof of \cref{lem:indep} we have
\begin{eqnarray}
\left\langle |\sum_k \alpha_k \beta_k^*|^2 \right\rangle & = & \int d\bm{\beta} \;\mathbb{P}_{\rm in}(\bm{\beta})
\sum_{k,l} \alpha_k \alpha_l |\beta_k|
|\beta_l|\exp(\i(\varphi_k-\varphi_l)) \nonumber \\
& =& \sum_{k,l} \alpha_k \alpha_l {\rm Tr} \prod_{j=0}^{J-1} \int d|\beta_j|  |\beta_j| |\beta_k| |
\beta_l| \int_{-\pi}^{\pi} d\varphi_j  \;
\; M_{\beta_j}^{\dagger} M_{\beta_j} \rho_{\rm in}\exp(\i(\varphi_k-\varphi_l))
\end{eqnarray}
with
\begin{equation*}
M_{\beta_j}^{\dagger} M_{\beta_j}=\frac{1}{\pi}\exp(-|\alpha_j|^2-|\beta_j|^2)\exp\left(K_{\beta_j}\cos(2\sqrt{\pi}\hat{q}-\varphi_j)\right)=\frac{1}{\pi}\exp(-|\alpha_j|^2-|\beta_j|^2)
\sum_{n_j \in \mathbb{Z}} I_{n_j}(K_{\beta_j}) S_q^{n_j}  \exp(-\i n_j \varphi_j).
\end{equation*}
When $k=l$, we see that the integrals over $\varphi_j$ lead to delta-functions at $n_j=0$ and the dependence on $\rho_{\rm in}$ drops out as we can use ${\rm Tr}\,\rho_{\rm in}=1$. For $k \neq l$, we project onto $n_{l}=-1$ and $n_k=+1$, picking up $I_{-1}(2\alpha_l |\beta_l|)S_q^{-1}$ and $I_{1}(2\alpha_k |\beta_k|)S_q$ factors. For $k\neq l$ we thus always apply a product $S_q S_q^{-1}=I$ and again the dependence on $\rho_{\rm in}$ drops out. Using that $I_{-1}(x)=I_1(x)$, we get

\begin{eqnarray}
\lefteqn{\left\langle |\sum_k \alpha_k \beta_k^*|^2 \right\rangle=} && \nonumber \\
& & 2^J \sum_{k=0}^{J-1} |\alpha_k|^2 \int d|\beta_k|  |\beta_k|^3  \exp(-|\alpha_k|^2-|\beta_k|^2) I_0(2 \alpha_k |\beta_k|) \times \prod_{j=0\colon j\neq k}^{J-1} \int d|\beta_j|  |\beta_j| \exp(-|\alpha_j|^2-|\beta_j|^2) I_0(2 \alpha_j |\beta_j|)  \nonumber \\
& & +\sum_{k\neq l=0}^{J-1} \alpha_k \alpha_l \int d|\beta_k|  |\beta_k|^2  \exp(-|\alpha_k|^2-|\beta_k|^2) I_1(2 \alpha_k |\beta_k|)\int d|\beta_l|  |\beta_l|^2  \exp(-\alpha_l^2-|\beta_l|^2) I_1(2 \alpha_l |\beta_l|)\nonumber \\
& & \times \prod_{j=0\colon j\neq k, j\neq l}^{J-1} \int d|\beta_j|  |\beta_j| \exp(-|\alpha_j|^2-|\beta_j|^2) I_0(2 \alpha_j |\beta_j|),
\end{eqnarray}
which can be simplified, using e.g. $\int_0^{\infty} dx \,x \exp(-x^2-y^2) I_0(2 y x)=1$, \cref{eq:fluc}, and $2 \int_0^{\infty} dx x^2 \exp(-y^2-x^2) I_1(2y x)=y$ to
\begin{eqnarray*}
\left\langle |\sum_k \alpha_k \beta_k^*|^2 \right\rangle & = &
\sum_k |\alpha_k|^2 (1+|\alpha_k|^2) +\sum_{k \neq l} |\alpha_k|^2 |\alpha_l|^2\\
& =& \sum_k |\alpha_k|^2+{\left(\sum_k |\alpha_k|^2\right)}^2  \\
& \rightarrow & |\alpha|^2 (1-\exp(-\kappa_{\rm open}t_{\rm meas}))(1+|\alpha|^2 (1-\exp(-\kappa_{\rm open}t_{\rm meas})))
\end{eqnarray*}
Thus when $t_{\rm meas}$ is long enough so that the entire state has leaked out, we can upper bound the expected $K_{\rm eff}\leq 2 |\alpha|\sqrt{1+|\alpha|^2}$, resulting in a lower bound on $\Delta_q$ equal to $1/\sqrt{4 \pi |\alpha| \sqrt{1+|\alpha|^2}}$. For long enough $t_{\rm meas}$ this is identical to our result for the direct measurement, which we have shown is closely related to the actual amount of squeezing in Fig.~\ref{fig:sq-result}.

\section{Details of Numerical Simulations}
The numerical simulations were implemented using the Qutip Python package.
In the numerics, we apply a counter-displacement drive $Z^{-\bar{n}}$, where $\bar{n}$ is the mean photon number of the initial state of the ancilla oscillator, in order to minimize the photon number of the state in the target oscillator. The time-evolution operator of the interaction between target and oscillator is then $U_{\rm PP}Z^{-\bar{n}}$, see the end of \cref{sec:circuit}.

\begin{figure}[htb]
	\includegraphics[width=.4\textwidth]{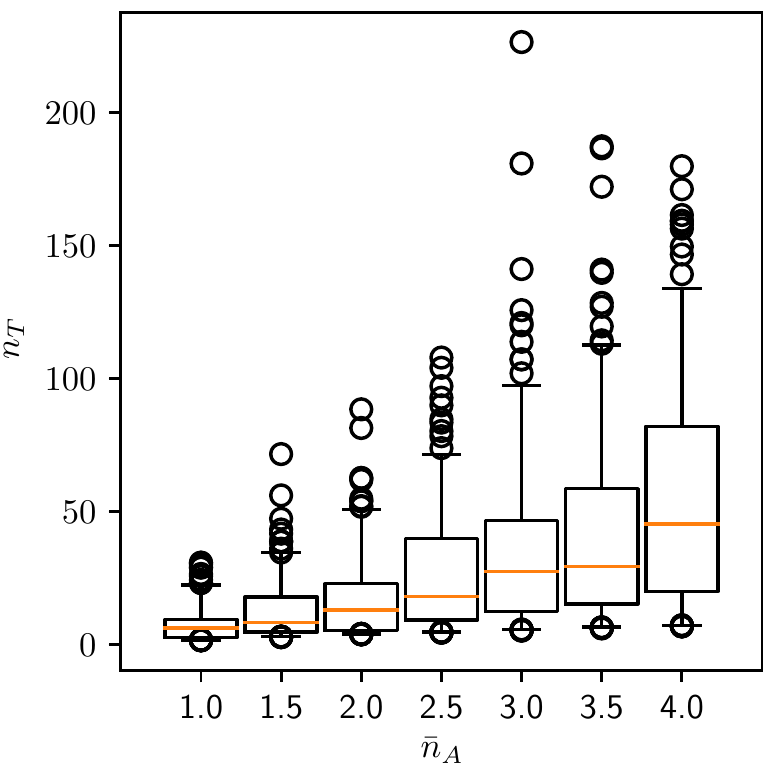}
	\hfill
	\includegraphics[width=.4\textwidth]{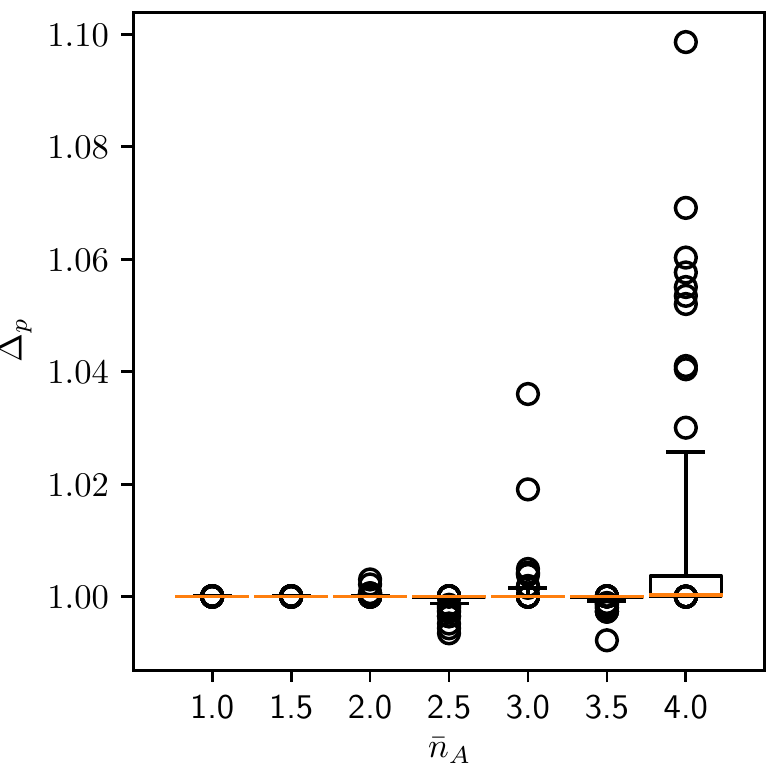}
	\caption{Box and whisker plot of the expected number of photons (left) and the effective squeezing $\Delta_p$ (right) of the final state after a measurement of $S_q$.
	The target oscillator starts in the vacuum state, and the ancilla oscillator with the coherent state $\ket{\alpha=\sqrt{\bar{n}}}$ (compare \cref{fig:sq-result}).
	For every $\alpha$, a total of 200 samples was simulated.
	The orange line indicates the median, the box indicates the 25 and 75 percentiles, the whiskers the 5 and 95 percentiles, events above or below these thresholds are shown individually.
	For some events, the final state of the target oscillator has a mean photon number exceeding 100, therefore a large Hilbert space is required to faithfully represent those states.
	From analytical considerations, we know that $\Delta_p=1$ should be constant, independent of the measurement results.
	As can be seen, errors are negligible up to $\bar{n}=3.5$, for $\bar{n}=4$, the relative error for most events is still below 1\%.}
	\label{fig:DeltaP_Vac}
\end{figure}

All simulations model a direct, perfect heterodyne measurement by projecting the ancilla oscillator onto a coherent state. The measurement result is chosen by sampling from the Husimi-Q function, using 200 randomly chosen samples unless mentioned otherwise.
Because this model of measurement is very strong, the photon numbers of the post measurement state may be very large, with some events exceeding 100 photons, see \cref{fig:DeltaP_Vac} on the left.
Note that for GKP states, the distribution of the photon number is very wide, with the standard deviation equal to the expected photon number.
Therefore, the Hilbert spaces of the target and ancilla oscillators were approximated using 500 and 20 Fock states, respectively.

To estimate the accuracy of the simulations, we use that the effective squeezing $\Delta_p$ of the vacuum state should stay constant in the case of a noiseless protocol.
The results are shown in \cref{fig:DeltaP_Vac} on the right. As shown there, errors are negligible up to an initial ancilla state with $\bar{n}=3.5$ photons, and the relative error for $\bar{n}=4$ photons is still below $1\%$ in most cases.
For these reasons, and because the effective squeezing achieved with $\ket{\alpha=2}$ as initial ancilla state is already very strong, we restrict the simulations to $\bar{n}=1,\dots,4$.

\bibliography{photonpress}
\end{document}